\documentclass[review]{elsarticle}
\usepackage{geometry}
\usepackage{lineno,hyperref}
\usepackage{dsfont}
\usepackage{tikz}
\usepackage{amsmath,amssymb,amsfonts,amsthm}

\usepackage{algorithm}
\usepackage{bm}
\usepackage{bookmark}
\usepackage{bm}
\modulolinenumbers[5]
\usepackage[all]{xy}
\usepackage{float}
\usepackage{extarrows}
\usepackage{graphicx} 
\usepackage{epstopdf}
\usepackage{subfigure}
\usepackage{gensymb}    
\usepackage{cases}
\usepackage[toc,page,title,titletoc,header]{appendix}

\usepackage{caption}
\usepackage{ulem}
\usepackage{cancel}
\usepackage{yhmath}
\usepackage{CJK}

\theoremstyle{definition}

\newdefinition{rmk}[thm]{Remark}

\theoremstyle{remark}

\journal{Journal of \LaTeX\ Templates}
\newcommand\addTwo[1]{{\color{black}{#1}}}
\newcommand\addZero[1]{{\color{black}{#1}}}
\newcommand\changeOne[1]{{\color{black}{#1}}}
\newcommand\changeTwo[1]{{\color{black}{#1}}}
\newcommand\add[1]{{\color{black}{#1}}}
\newcommand\del[1]{}
\newcommand\delEq[1]{}

\makeatletter
\def\ps@pprintTitle{%
   \let\@oddhead\@empty
   \let\@evenhead\@empty
   \def\@oddfoot{\reset@font\hfil\thepage\hfil}
   \let\@evenfoot\@oddfoot
}
\makeatother

\begin{document}
\pdfoutput=1\begin{CJK*}{UTF8}{gbsn}
\pdfoutput=1\begin{frontmatter}
\pdfoutput=1\title{Self-propulsion dynamics of small droplets on general surfaces with curvature gradient}

\pdfoutput=1\author[]{Yujuan Chen}
\pdfoutput=1\ead{cyj@lsec.ac.cc.cn}
\author[]{Xianmin Xu\corref{cor1}}
\ead{xmxu@lsec.ac.cc.cn}

\address{LSEC, Institute of Computational Mathematics and Scientific/Engineering Computing, NCMIS, AMSS, Chinese Academy of Sciences, Beijing 100190, China}
\address{School of Mathematical Sciences, University of Chinese Academy of Sciences, Beijing 100049, China}
\cortext[cor1]{The author to whom correspondence may be addressed:  xmxu@lsec.cc.ac.cn.}

\fntext[fn1]{The work was partially supported by NSFC 11971469 and by the National Key R\&D Program of China under Grant 2018YFB0704304 and Grant 2018YFB0704300.\;}


%

\begin{abstract}
We study theoretically the self-propulsion dynamics of
a small droplet  on general curved surfaces by a variational approach.
A new reduced model is derived based on careful computations for the capillary energy and the viscous dissipation in the system.
The model describes quantitatively the spontaneous motion of a liquid droplet on general surfaces.
In particular, it recovers previous models for droplet motion on the outside surface of a cone.
In this case, we derive a  scaling law of the displacement $s\sim t^{1/3}$ of a droplet with respect to time $t$ by asymptotic analysis.
Theoretical results are in good agreement with  experiments in previous literature
without adjusting
the friction coefficient in the model.
\end{abstract}

\end{frontmatter}
\end{CJK*}

\section{Introduction}
Self-propulsion of droplets on solid substrates  is common in nature and  has been widely exploited  by plants and animals for different purposes, for example,  fog collection  by cactus spine\cite{Ju2012A} and spider silk\cite{zheng2010directional},  water separation from the legs of water striders\cite{Qianbin2015Self}, and many others\cite{prakash2008surface,duprat2012wetting,liu2015effective}.
The  phenomenon also has important applications in many industrial processes,
such as heat transfer, microfluid devices, water-repellent surfaces, etc. Therefore,
the problem have been studied extensively in  experiments and theory(see \cite{chaudhury1992make,dos1995free,lorenceau2004drops,duprat2012wetting,mayo2015simulating,chen2018ultrafast,wang2020enhancing}
and literatures therein).

The directional transportation of droplets can be induced by various properties of the substrates.
One common reason for self-propulsion is  the wetting gradient, which can be caused by either
the chemical inhomogeneity of the surface \cite{sumino2005self, sumino2005chemosensitive,zheng2010directional,xu2012droplet,li2017external}
or the  carefully designed roughness of the substrate\cite{yang2008conversion}. In addition,
temperature gradient~\cite{xu2012thermal,bjelobrk2016thermocapillary} and external fields\cite{li2017external} may
also cause wetting gradient.
On such surfaces, a droplet moves from a more  hydrophobic part to a more hydrophilic region to minimize the total wetting energy
in the system.
The directional motion of a droplet can also be caused by  geometric properties  of the substrates, e.g.
the varying distance between two plates  in a wedge region~\cite{prakash2008surface,reyssat2014drops,wang2020enhancing} or
 the curvature gradient on a conical surface\cite{lorenceau2004drops,renvoise2009drop,chan2020directional}.

Recently, the motion of a small droplet on a conical surface has arisen much interest \cite{LvSubstrate,galatola2018spontaneous,JohnDynamics}.
In \cite{LvSubstrate}, Lv {\it et al.} presented the first experiment for
the spontaneous motion of a water droplet on conical surfaces.
 They also computed the driving capillary force \del{in a clever way} by
computing the surface energy of a droplet on a spherical substrate
 to approximate the energy on  the conical surface. 
Later on, the problem was studied theoretically by Galatola \del{in} \cite{galatola2018spontaneous}.
\del{He}\addZero{The author} computed the capillary force by a perturbative analysis for the surface of a droplet in equilibrium.
The force  is balanced by a viscous friction force
of the contact line and this leads to a reduced model.
The model generates results in agreement with the experiments in~\cite{LvSubstrate} when
the friction coefficient is adjusted as a fitting parameter.
More recently, McCarthy {\it et al.} studied the dynamics of a droplet on a conical surface covered with silicone oil both in experiments and
in theory\cite{JohnDynamics}. In their work, the energy dissipation in the oil film is dominant and the viscous dissipation in the droplet is ignored.
{\color{black}The main drawback of the previous theoretical analysis is that the viscous dissipations in
the droplet have not been considered.
For flow with moving contact lines, it is usually very challenging to  compute the energy dissipations in
the system. There
is no analytical solution even for a sliding droplet problem in steady state on planar surfaces.
In addition, most of the previous studies are for droplet motion on conical surfaces.
 There seems few work for the general surface cases, which might be useful in industrial applications.
}

In this work, we study {\color{black}theoretically} the self-propulsion transportation of a small droplet on  a general surface with curvature gradient.
We derive a new reduced model for the droplet motion by using the Onsager principle as an approximation tool (\cite{onsager1931phys, onsager1931phys1,doi2013soft}). The method is  equivalent to directly balance
the driving force and the friction force in the system~\cite{galatola2018spontaneous}.
We exploit the idea in \cite{LvSubstrate} to compute approximately the total energy
on a general surface. For the  dissipation function, we carefully compute the viscous dissipation in
the droplet, which leads to an effective viscous friction of the contact line.
{\color{black} More precisely, we separate the droplet into three subregions,
namely a bulk region far from the contact line, a microscopic region with nano-scale in the vicinity of the contact line
and a mesoscopic wedge region in between.
 By careful analysis and numerical simulations, we show that the dissipations in the wedge region are
 dominate. Therefore, we consider only the viscous dissipations in the wedge region in our theory.}
 The reduced model is an ordinary differential
equation intrinsically defined on the surface. The direction of the droplet motion is along the surface gradient
direction of the mean curvature of the substrate.
 Numerical examples show that the model
can be used to simulate the spontaneous transport of a droplet on general surfaces.
We also show that our theory recovers formally the model in~\cite{galatola2018spontaneous} in the conical surface case: the
driving capillary force is consistent with that in~\cite{galatola2018spontaneous} in leading order while
the effective viscous friction coefficient is given explicitly and is not a fitting parameter any more.
The model can be further improved by adding an acceleration term when the inertial effect can not be ignored.
Numerical results by our model  fit well with the
experimental data in \cite{LvSubstrate} even without adjusting the friction coefficient.
 In addition, we show that the displacement $s$ of the droplet on the outer conical surface has a scaling law  $s\sim t^{1/3}$
 for relatively late time. This scaling law is different from that  in~\cite{JohnDynamics}, where
the energy dissipation is dominate in the oil film coated on the substrate and that leads to a slower power law  $s\sim t^{1/4}$.



The rest of the paper is organized as follows. In section 2, we derive a reduced model for droplet motion on general surfaces by
the Onsager variational principle. Both the driving force and the viscous dissipations are computed approximately.
In section 3, we consider
a conical surface case to compare with previous theoretical and experimental work. We derive a scaling law of the displacement of
the droplet on the outer surface of the cone. In section 4, we present some numerical results for both general surfaces and conical surfaces.
The results on conical surfaces are in good agreement with previous experiments
without adjusting the friction coefficient. Some conclusion remarks are given in the last section.

\section{Variational analysis for droplet motion on a curved surface}
\begin{figure}[H]
  \centering
  \includegraphics[width=2.5in]{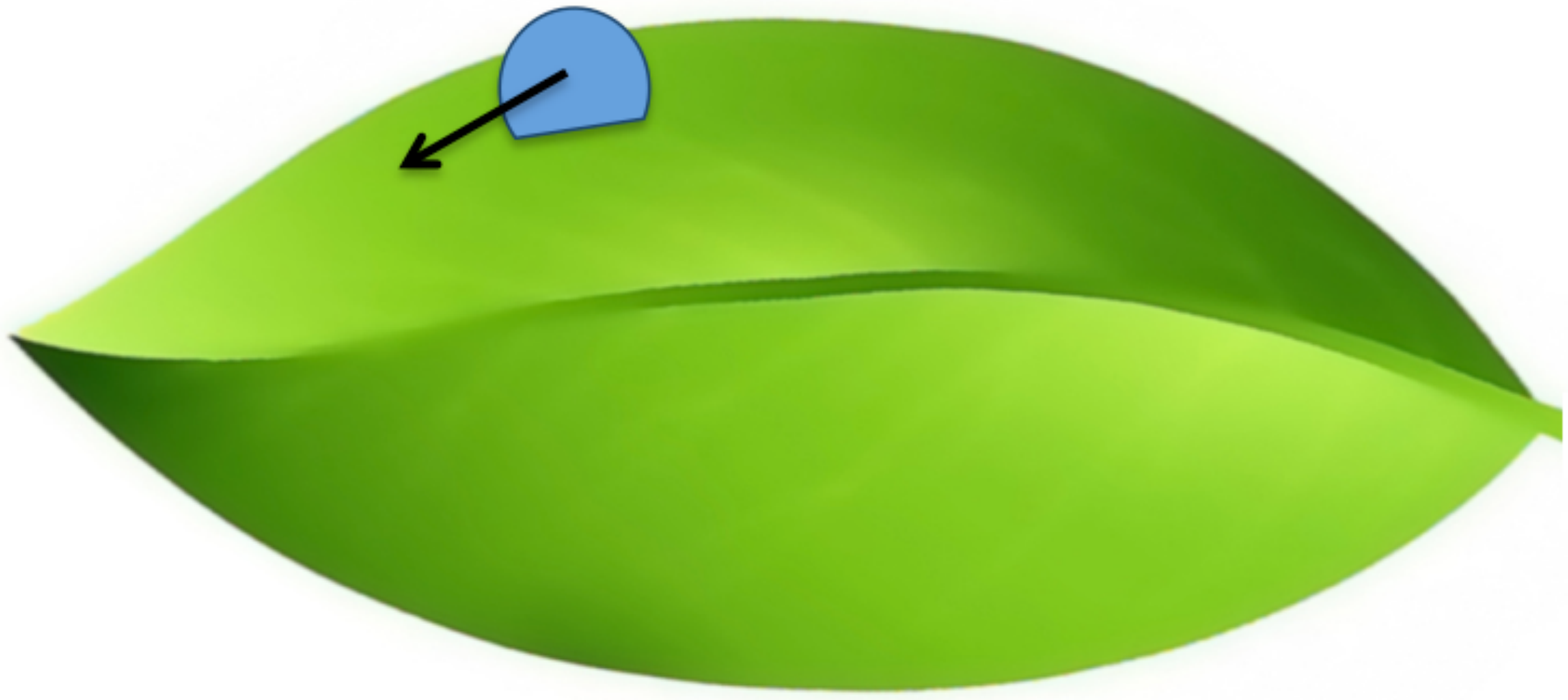}\\
  \caption{A droplet on a general surface with curvature gradient}\label{gs}
\end{figure}

We consider a droplet motion problem on a general surface $\Gamma$ as shown in Fig.~\ref{gs}.
The volume of the droplet is denoted $V_0$. We assume that
the characteristic size of the droplet is smaller than the capillary length $L_c=\sqrt{\gamma/\rho g}$,
where $\gamma$ is the surface tension of the liquid surface,
$\rho$ is the density of the liquid and $g$ is the gravitational acceleration, so that
 the effect of gravity can be ignored.
On the general surface, the equilibrium contact angle $\theta_e$ of the bubble is given by the Young's equation,
\begin{equation}\label{e:Young}
\gamma\cos\theta_e=\gamma_{SV}-\gamma_{SL},
\end{equation}
where $\gamma_{SV}$ and $\gamma_{SL}$ are respectively surface tensions of the solid-air interface
and the solid-liquid interface.

We will use the Onsager principle as an approximation tool to study
the motion of the droplet on the surface.
The Onsager principle is a variational principle that the Rayleighian $R(\dot{x},x)=\Phi(\dot{x},x)+\dot{\mathcal{E}}(x)$ of a system is minimized with respect to the changing rate $\dot{x}(t)$ for a set of parameters  $x(t)=\{x_1(t),x_2(t)\ldots,x_N(t)\}$.
Here $\Phi(\dot{x},x)$ is the energy dissipation function which is defined as the half of  the energy dissipated per unit time
 and $\mathcal E(x)$ is the total energy of the system.  The parameter set $x(t)$ can be
a set of slow variables to characterize the system.
Then the principle leads to a reduced model
$$\frac{\partial \Phi}{\partial \dot{x}_i}+\frac{\partial \mathcal{E}}{\partial x_i}=0,$$
 which represents the balance of the friction  force $\frac{\partial \Phi}{\partial \dot{x}_i}$ and the driven force $-\frac{\partial \mathcal{E}}{\partial x_i}$.
The method has been used to derive approximate models  in many
  problems in soft matter and in hydrodynamics
 (see e.g. in \cite{Doi15,xu2016variational,guo2019onset,ManDoi2016,doi2019application,xu2020theoretical}). In particular,
it was used to study solid dewetting on curved substrates in \cite{jiang2018solid}.

To use the Onsager principle, we need specify some slow variables by introducing some {\it ansatz}
on the shape of the droplet. As shown by the recent analysis by Galatola\cite{galatola2018spontaneous},
the shape of the droplet is just a small perturbation of a spherical surface. In leading order approximation,
we can simply  assume that  the droplet is spherical. We also assume that the contact angle of the droplet
with the substrate is almost equal to the equilibrium contact angle $\theta_e$. Then the shape of the droplet can be determined once the local
curvature of the substrate is known.
We are interested in how the droplet moves on the curved solid surfaces.
We  denote by $\vec{x}$  the position of the droplet on the substrate $\Gamma$
and use it as the slow variable to characterize the system.

In the following, we will first compute approximately the total free energy of the droplet on the curved surface
and then compute the viscous energy dissipation in the system.
Finally we will use the Onsager principle to derive the dynamic
equation of the droplet motion on the surface.
\subsection{The surface energy}
To compute the total free energy of the droplet on a curved surface,
we use the idea as in Lv {\it et~al} \cite{LvSubstrate}. Suppose the droplet is located at a point $\vec{x}\in\Gamma$ at time $t$.
The mean curvature of the surface on this point is given by $\kappa(\vec{x})$.
\addTwo{Let $r$ represent the characteristic size of the droplet. We assume the droplet is small in the sense that
$r|\kappa(\vec{x})|\ll 1$.
By the  theory of differential geometry, it is known that
the deviation of the substrate from its tangent plane  at $\vec{x}$ is determined
by the two principal curvatures $\kappa_1$ and $\kappa_2$ of the surface at this point.
We write the total interface energy of the droplet as a function of $r \kappa_1$ and $r\kappa_2$, i.e. $\mathcal{E}(r\kappa_1, r\kappa_2)$.
Suppose the liquid surface is isotropic,  then we have
 \changeTwo{$\frac{\partial \mathcal{E}}{\partial(r \kappa_1)}\Big|_{(0,0)} =\frac{\partial \mathcal{E}}{\partial(r \kappa_2)}\Big|_{(0,0)}$}. The Taylor expansion of $\mathcal{E}(r \kappa_1, r \kappa_2)$ is expressed as(see Supplemental Material in \cite{LvSubstrate}):
\changeTwo{
\begin{align}
  \mathcal{E}(r \kappa_1, r \kappa_2) &= \mathcal{E}(0,0)+
  \frac{\partial \mathcal{E}}{\partial(r \kappa_1)}\Big|_{(0,0)}r \kappa_1
  +\frac{\partial \mathcal{E}}{\partial(r \kappa_2)}\Big|_{(0,0)} r \kappa_2 +O((r\kappa)^2)\nonumber\\
 & = \mathcal{E}(0,0) + 2\frac{\partial \mathcal{E}}{\partial(r \kappa_1)}\Big|_{(0,0)}r \kappa +O((r\kappa)^2),\label{e:temp}
\end{align}
}
where $\kappa=(\kappa_1+\kappa_2)/2$ is the mean curvature.
This means that if we ignore the higher order terms,  $\mathcal{E}$ only depends on the mean curvature but not on the Guass curvature $\kappa_1\kappa_2$.
Therefore we can assume that the shape of the droplet is approximated well by
its equilibrium state on a spherical surface with the same mean curvature.} The validity of the approximation is verified by perturbative analysis in \cite{galatola2018spontaneous,JohnDynamics}. \addTwo{ In addition,  comparisons with the exact numerical solutions for
a droplet on conic surfaces show that the approximation is still quite accurate when $r_c|\kappa|\leq 0.2$,
where $r_c$ is the averaged radius of the contact line between the droplet surface and the substrate\cite{galatola2018spontaneous}.}

As mentioned above, we can ignore the gravity of the droplets and  the total free energy is given by
$\gamma A_{LV}+\gamma_{SL}A_{SL}+\gamma_{SV}A_{SV}$,
where $A_{LV}$, $A_{SL}$ and $A_{SV}$ are respectively the areas of liquid-gas, solid-liquid and solid-gas interfaces.
By diminishing a constant $\gamma_{SV}(A_{SL}+A_{SV})$, the total energy is rewritten as:
\begin{equation}
\begin{aligned}
\label{UUU}
\mathcal{E}=\gamma(A_{LV}-A_{SL}\cos \theta_e),
\end{aligned}
\end{equation}
where we have used the Young's equation~\eqref{e:Young}.


 \begin{figure}[H]
 \subfigure
 {
 \begin{minipage}[t]{0.5\linewidth}
 \centering
 \includegraphics[width=2in]{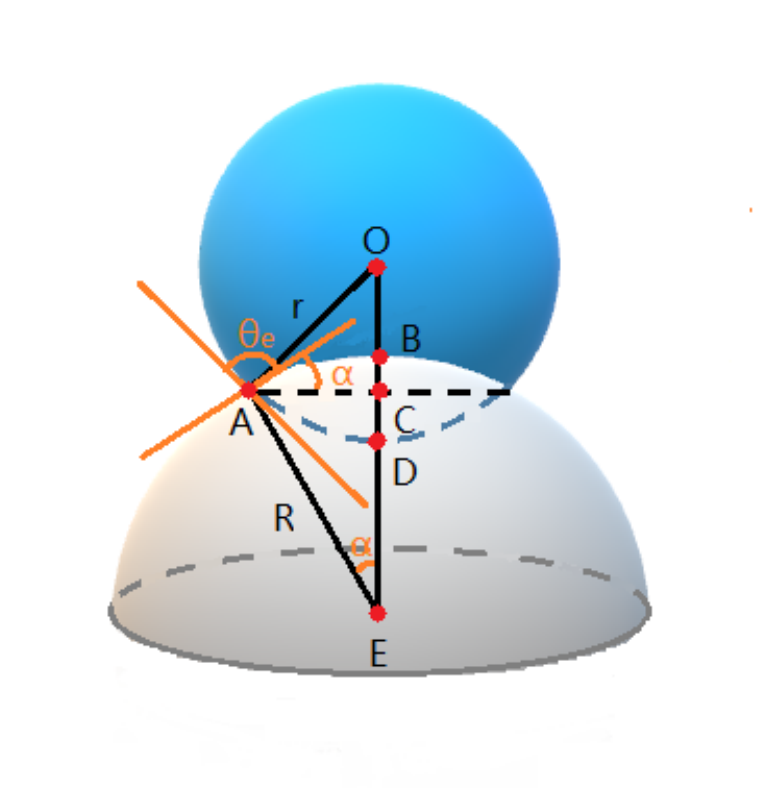}
 \end{minipage}%
 }
 \subfigure
 {
 \begin{minipage}[t]{0.5\linewidth}
 \centering
 \includegraphics[width=2in]{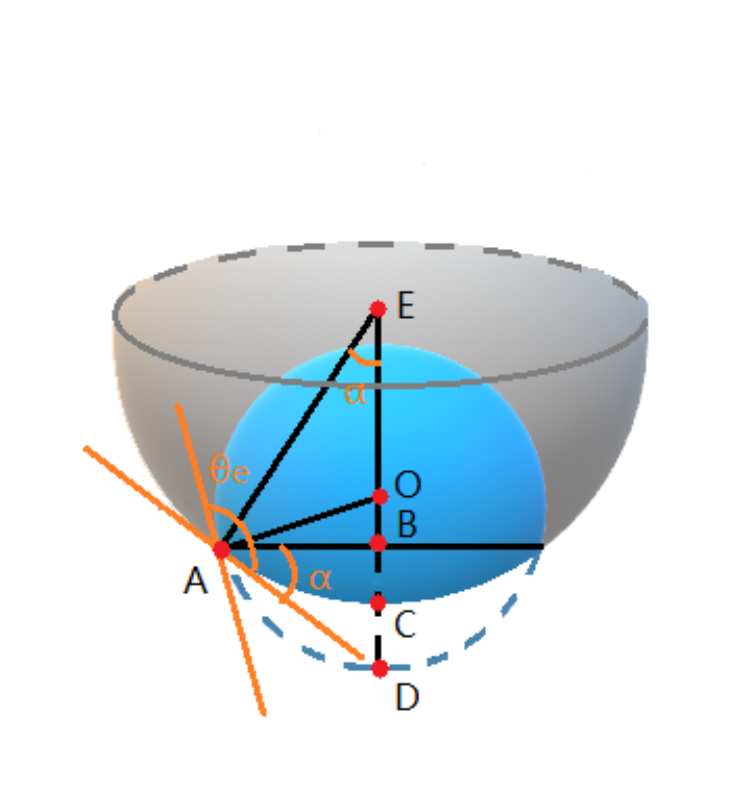}
 \end{minipage}
 }
  \caption{A droplet placed on the outside and inside surfaces of a larger sphere. Left: The negative mean curvature case($\kappa<0$);
  Right: The positive mean curvature case($\kappa>0$).}\label{0}
\end{figure}

 We first compute the total surface energy of a stationary droplet on a spherical surface as shown in Fig~\ref{0}.
On the surface, the stationary state of the droplet  will  take spherical shape with a Young's angle $\theta_e$.
When the volume $V_0$ and the Young's contact angle $\theta_e$ are given, the free energy
$\mathcal{E}$ of the droplet can be written a function of the radius $R$ of the substrate.
The calculation is as follows. We first consider the case when the droplet is put on the outside surface of the sphere.
We denote by $r$ the radius of the droplet and by $\alpha$ the cap angle  of the solid-liquid surface(see Fig.~\ref{0}(a)). 
 Then we can calculate the total free energy $\mathcal{E}$ by the Eq.~\eqref{UUU}:
\begin{equation*}
\begin{aligned}
\label{U}
\mathcal{E}=2\pi \gamma \left[\displaystyle r^2(1-\cos\hat{\theta})- R^2(1-\cos\alpha)\cos \theta_e \right],
\end{aligned}
\end{equation*}
where $\hat{\theta}=\theta_e+\alpha$. The volume of the droplet is given by
\begin{equation*}
  V_0=\displaystyle\frac{\pi}{3}r^3 (1-\cos \hat{\theta})^2(2+\cos \hat{\theta})-\frac{\pi}{3}R^3(1-\cos\alpha)^2(2+\cos\alpha). \label{V1}
\end{equation*}
By the geometric constraint
\begin{equation*}
R\sin\alpha =r \sin \hat{\theta}, \label{V2}
\end{equation*}
we have $r=R \sin\alpha/\sin\hat{\theta}$.
\del{We denote $a=\sin\alpha$. Then both $\mathcal{E}$ and $V_0$  can
be written as a function of $R$ and $a$, that}
\addZero{Then both $\mathcal{E}$ and $V_0$  can
be written as a function of $R$ and $\alpha$, that
\begin{equation}
\begin{aligned}
\label{eq:U}
\mathcal{E}=2\pi R^2\gamma \left[\displaystyle\frac{ \sin^2\alpha}{1+ \cos\hat{\theta} }- (1-\cos\alpha) \cos \theta_e \right],
\end{aligned}
\end{equation}
and
\begin{equation}
\begin{aligned}
\label{V}
V_0=\displaystyle\frac{\pi}{3}R^3 \left[\sin^3\alpha \frac{(1-\cos\hat{\theta})(2+\cos\hat{\theta})}{\sin\hat{\theta}(1+\cos\hat{\theta})}- (2-3\cos\alpha+\cos^3\alpha)\right].
\end{aligned}
\end{equation}
}
For any given $V_0$, \del{$a$}\addZero{$\alpha\in [0,\pi]$} is implicitly determined by $R$ from \eqref{V} and the energy $\mathcal{E}$
can be written as a function of $R$.
The same results can be obtained when the liquid droplet is placed on the inside surface of the sphere (see Fig.~\ref{0}(b)),
only noticing that $\alpha$\del{(and accordingly $a$)} is negative in this case.

On a general surface $\Gamma$, its mean curvature $\kappa(\vec{x})$ only depends on
the position $\vec{x}=(x_1,x_2,x_3)$ of the surface. If a small droplet (with volume $V_0$) is put on the surface,
 its shape is approximated well by a stationary droplet on a spherical substrate with  the same mean curvature.
 We assume the mean curvature of the outer surface of a sphere is negative, so that $\kappa=-{R}^{-1}$.
Then the total energy on a substrate with a general shape can be approximated by
\addZero{
\begin{align}
\mathcal{E}&=2\pi (-\kappa)^{-2}\gamma \left[\displaystyle\frac{ \sin^2\alpha}{1+ \cos\hat{\theta} }- (1-\cos\alpha) \cos \theta_e \right],\label{eq:generalU}\\
V_0&=\displaystyle\frac{\pi}{3}(-\kappa)^{-3} \left[\sin^3\alpha \frac{(1-\cos\hat{\theta})(2+\cos\hat{\theta})}{\sin\hat{\theta}(1+\cos\hat{\theta})}- (2-3\cos\alpha+\cos^3\alpha)\right].\label{generalV0}
\end{align}
}
Once again \del{$a$}\addZero{$\alpha$} can be written as a function of $\kappa$ by the equation~\eqref{generalV0} for a given volume $V_0$. Then the first equation~\eqref{eq:generalU}
implies that $\mathcal{E}$ is a function of $\kappa$. Notice that $\kappa$ depends only on $\vec{x}$,
the total energy $\mathcal{E}$ can be regarded as a function of the position of the droplet.

Suppose the location of the droplet is $\vec{x}\in\Gamma$, then we can compute
derivative of the energy $\mathcal{E}$ with respect to $\vec{x}$,
\begin{equation}\label{e:Edx}
\frac{\delta \mathcal{E}}{\delta \vec{x}}=\frac{d\mathcal{E}}{d \kappa}\nabla_{\Gamma} \kappa(\vec{x}),
\end{equation}
 where $\nabla_\Gamma \kappa(\vec{x})$ is the surface  gradient of $\kappa$ on the substrate. Using \eqref{eq:generalU} and \eqref{generalV0},
direct calculations on the derivative of $\mathcal{E}$ with respect to $\kappa$ give
\del{where $X$ and $Y$ are given in \eqref{X} and \eqref{Y}, and} 
\addZero{
\begin{equation}
\begin{aligned}
\label{generaldEdk}
\frac{d \mathcal{E}}{d \kappa}=\frac{\partial \mathcal{E}}{\partial \alpha}\frac{d \alpha}{d \kappa}+\frac{\partial \mathcal{E}}{\partial \kappa}
=&\frac{6\gamma V_0}{A} \left[\frac{-2\cos\alpha }{1+\cos\hat{\theta}}-\frac{\sin\alpha \sin\hat{\theta}}{(1+\cos\theta)^2}+\cos\theta_e\right]\\
&+4\pi\gamma (-\kappa)^{-3}\left[\frac{\sin^2\alpha}{1+\cos\hat{\theta}}-(1-\cos\alpha)\cos\theta_e\right],
\end{aligned}
\end{equation}
where $A$ is given by 
\begin{equation}
\begin{aligned}
\label{A}
A=&\sin\alpha \cos\alpha \frac{(1-\cos\hat{\theta})(2+\cos\hat{\theta})}{\sin\hat{\theta}(1+\cos\hat{\theta})}+\sin^2\alpha \left[\frac{1}{(1+\cos\hat{\theta})^2}-1\right] .
\end{aligned}
\end{equation}
}
Notice that the right hand side term of \eqref{generaldEdk} represents a complicated function of $\kappa$ since  \del{$a$}\addZero{$\alpha$} is an
implicit function of $\kappa$ by \eqref{generalV0}. Here $\theta_e$ and $V_0$ are given parameters.

\addTwo{We can simplify the formula for the surface energy $\mathcal{E}$ and its derivative by asymptotic
analysis when the droplet is small as in Appendix B.
In this case, $a=\sin\alpha$ can be chosen as a small parameter.
By ignoring some higher order terms, the total energy $\mathcal{E}$ is approximated by (see Eq.~(\ref{appro_E}))
\begin{equation}\label{e:approxE}
\mathcal{E}\approx \gamma \sin {\theta_e}\left[(3V_0)^{\frac{2}{3}}(\pi Z_1)^{\frac{1}{3}} -\frac{3V_0 \kappa}{2Z_1}  \right].
\end{equation}
where $Z_1=\frac{1}{\sin\theta_e}\left(\frac{2}{1+\cos\theta_e}-\cos\theta_e\right)=\frac{\sin\theta_e(2+\cos\theta_e)}{4\cos^4(\theta_e/2)}
$.
 The  analysis for \eqref{generaldEdk}  in~ Appendix B also shows that
\begin{equation}\label{e:approxDE}
\frac{d \mathcal{E}}{d \kappa }\approx -\frac{3\gamma V_0\sin\theta_e}{2Z_1}=-\frac{6 \gamma V_0 \cos^4(\theta_e/2) }{2+\cos\theta_e}.
\end{equation}
It is equivalent to the direct calculation for the derivative of the approximate energy in~\eqref{e:approxE}.
}

\subsection{The energy dissipation}
\begin{figure}[H]
  \centering
  \includegraphics[width=4.5in]{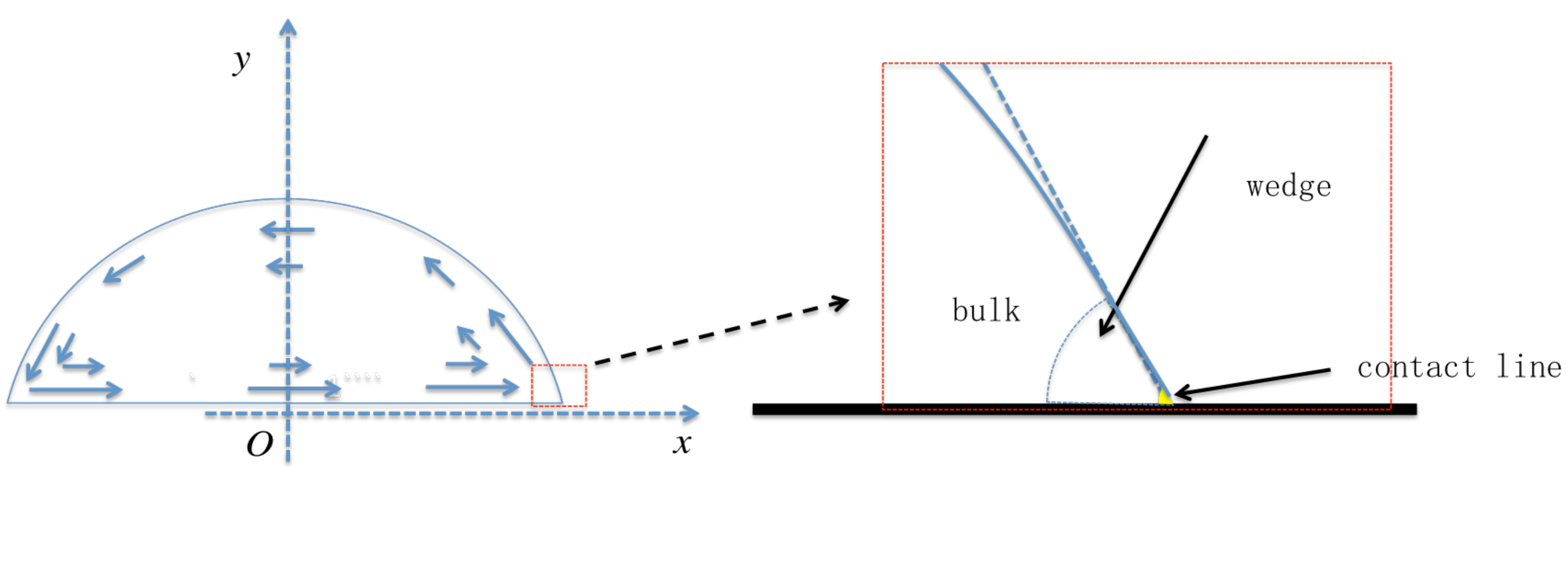}
  \vspace{-0.5cm}
  \caption{Different regions in a liquid drop. Left: Velocity field in the droplet
      in  a frame moving with the droplet. Right: The contact line region.
  \add{A velocity field by direct simulations is shown in Fig.~\ref{fig:vel} in Appendix A}}\label{fig:wedge}
\end{figure}

We then compute the viscous energy dissipation in the droplet.
This is a typical moving contact line problem.
As shown in \cite{cox1986,bonn2009wetting}, we separate the system into three different regions(see Figure~\ref{fig:wedge}).
In the vicinity of the contact line with distance smaller than $l_c\sim 1 \mathrm{nm}$, the molecular interactions between the
liquid and solid are important. In this region, the energy dissipation can be written in a formula corresponding to a friction force
of the contact line\cite{guo2013direct}. The immediate region  can be regarded as a wedge region of a mesoscopic size $l$ near the contact line,
where the  classical hydrodynamics are applicable. The rest is the outer  bulk region, where
the velocity field of the droplet is in macroscopic scale.
For a small droplet, the viscous dissipation in the bulk region is less important than that in the wedge region near the contact line when the
capillary number \add{$Ca={\eta v}/{\gamma}$} is small\cite{Gennes03,bonn2009wetting}, \add{
where $\eta$ is the  liquid viscosity and $v$ is the characteristic velocity}.  
The fact has also been used a lot in literature(c.f.\cite{tanner1979spreading,Gennes03,XuWang2020}), including the recent analysis for a  droplet with a barrel configuration which engulfs a portion of a conical fiber\cite{chan2020directional}.
In the following, we only consider the energy dissipation in the first two regions.

We first consider the dissipation in the vicinity of the contact line. The dissipation is due to the contact line friction.
Denoted by $v_{ct}$ the  velocity of a contact line. Then the local dissipation rate per unit length of the contact line can
be written approximately as $\Psi_0={\xi_0} v_{ct}^2$, where $\xi_0$ is the contact line viscosity coefficient(as used in \cite{galatola2018spontaneous}).
 The  coefficient $\xi_0$ has been
directly measured   by experiments in \citep{guo2013direct}. It is found that $\xi_0$ can be approximated well by
$(0.8\pm 0.2) \eta$. \del{where $\eta$ is the fluid viscosity.}

We then compute the dissipation in the mesoscopic wedge region near the
contact line with apparent contact angle $\theta_e$. 
When the contact angle $\theta_e$ is small, the viscous dissipation in a two-dimensional  wedge region is approximated by
the well-known formula (\cite{Gennes03})
\begin{equation}\label{e:dissipW0}
\Psi \approx \frac{3\eta|\ln \varepsilon|}{\theta_e} v_{ct}^2,
\end{equation}
 where the dimensionless parameter $\varepsilon=l/l_{c}$ is the ratio between the mesoscopic size $l$ and the microscopic size $l_c$, $v_{ct}$ is the velocity of the contact line.
When the contact angle $ \theta_e $ is large, the viscous dissipation in the wedge can be computed by solving
a Stokes equation with specified boundary condition(see in Appendix A and also in \cite{XuWang2020}).  
The  energy dissipation rate in the two-dimensional wedge region is given by
\begin{equation}\label{e:dissipW1}
  \Psi \approx \frac{2 \eta\sin^2\theta_e|\ln\varepsilon| }{\theta_e-\sin\theta_e \cos\theta_e}v_{ct}^2:=\xi_1(\theta_e,\eta,\varepsilon) v_{ct}^2
\end{equation}
One can easily verify that the formula approaches to that in~\eqref{e:dissipW0} when $\theta_e$ is small.
From~\eqref{e:dissipW1}, we could see that the dissipation is a quadratic form with respect to the contact line velocity with
an effective friction coefficient $\xi_1$ depending on $\eta$, $\theta_e$ and the cut-off parameter $\varepsilon$. In general, the term $\ln\varepsilon$ is  $O(10)$(see~\cite{Gennes03}). {\color{black} For example, if we choose $l_c \approx 1 \mathrm{nm}$ and $l\approx 50 \mu \mathrm{m}$, the we have $\ln\varepsilon\approx 10.82$. } Then simple computations show that $\xi_1$ is at least ten times larger than $\xi_0$.
This indicates that the viscous dissipation in the wedge  region is dominant with respect to the friction dissipation of the contact line.

\begin{figure}[H]
  \centering
  \includegraphics[width=3.in]{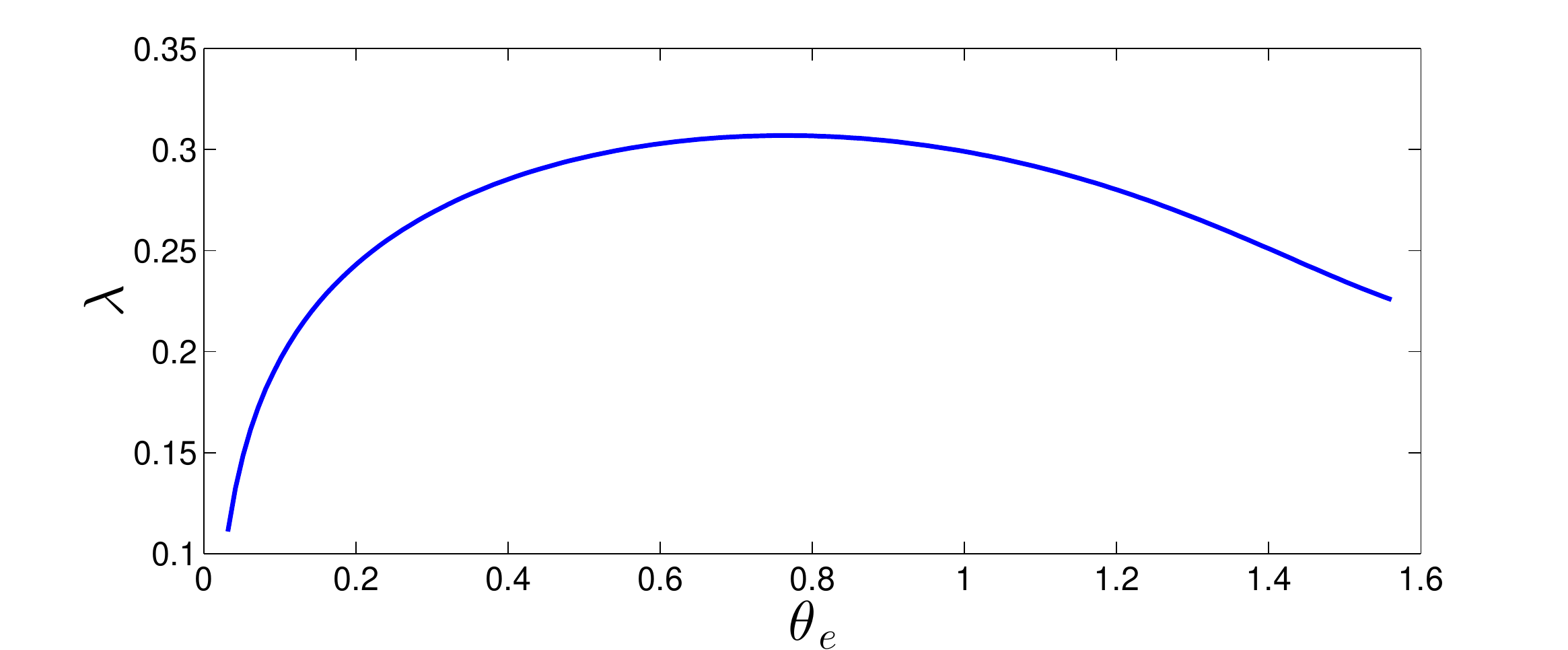}\\
  \caption{Ratio of the dissipation coefficients for the bulk and wedge regions(in two dimensions). }\label{fig:dissp}
\end{figure}

{\color{black} We now consider the viscous energy dissipations in the bulk region. In this case, we need to solve a Stokes equation in a  region with a spherical
surface. In general, the problem can not be solved analytically, since the velocity field can not be axisymmetric.
 Here we give some estimate for the viscous dissipations in the bulk region. For simplicity, we consider a two dimensional
problem as shown in Figure~\ref{fig:wedge}. 
In a frame moving with the droplet, we can suppose
the horizontal velocity on the substrate is $U$. Denote by $h(x)$ the local height of the droplet.
The shear stain is bounded by $c_0 U/h(x)$ where $c_0$ is a constant. Suppose the radius of droplet is $R$ and the contact angle is $\theta_e$.
Then we have $h(x)=\sqrt{R^2-x^2}-R\cos\theta_e$.
  The bulk dissipation can be bounded by
 $$
 c_0^2\eta\int_{0}^{R\sin\theta_e-l}  \frac{U^2}{h(x)} dx=  c_0^2\eta U^2 \int_{0}^{\sin\theta_e-l/R}  \frac{1}{\sqrt{1-\hat{x}^2}-\cos\theta_e} d\hat{x}=c_0^2\eta U^2 f(\theta_e, l/R),
 $$
 where $f(\theta_e,l/R)= \int_{0}^{\sin\theta_e-l/R}  ({\sqrt{1-\hat{x}^2}-\cos\theta_e})^{-1} d\hat{x}$ can be calculated numerically.
In our cases, $l\approx 50\mu \mathrm m$ and $R\approx 1 \mathrm{mm}$ so that $l/R\approx 0.05$.
In steady state, the velocity $U$ is the same as the velocity $v_{ct}$ of the contact line.
To compare the bulk dissipations with  those in the wedge region,
  we could compute the ratio $\lambda(\theta_e)= \eta c_0^2 f(\theta_e,0.05)/\xi_1(\theta_e,\eta,\varepsilon)=c_0^2 f(\theta_e,0.05)(\theta_e-\sin\theta_e \cos\theta_e)/2 \sin^2\theta_e|\ln\varepsilon|  $. Some typical results are
 shown in Figure~\ref{fig:dissp} where we choose $c_0^2=3$ and $\ln\varepsilon=10.82$.
 We could see that the ratio 
  ranges from
  $0.1$ to $0.3$ for the two-dimensional problem.
 We also verify this by solving a Stokes equation numerically in a circular domain.
 The typical values for the ratio are around $0.22$(see Appendix A).
 Notice that the ratio between the bulk dissipations and those near the contact line
  should be smaller than that in two dimensions.
Therefore, we could conclude that the viscous dissipation in the bulk is less important than that in the mesoscopic wedge region
in our cases.}
 {In the following analysis, we consider only the viscous dissipation in the mesoscopic
 wedge region near the contact line. 
}

{
We now compute the energy dissipation function $\Phi$ of the moving droplet on the substrate, which is defined as half of the total viscous dissipation.
We suppose that the contact line on the conical surface can be approximated well by a circle with radius
$r_c= (-\kappa)^{-1} \sin\alpha$
(see in Figure~\ref{0}).  Notice that $ (-\kappa)^{-1} \sin\alpha$ is always positive
since the mean curvature $\kappa$ of the substrate and $\sin\alpha$ always have opposite signs. 
Then the energy dissipation function is calculated by integration of ${\Psi}/{2}$ along the contact line,
\begin{equation}\label{e:Phi}
\begin{aligned}
  \Phi &= \frac{1}{2}\int_{CL}\xi_1 v_{ct}^2 dS= \frac{ \delEq{(-\kappa)^{-1}(\sin\alpha)}\addZero{r_c}\xi_1}{2}\int_0^{2\pi} |\vec{v}_{ct}|^{2}  d\theta  
  =\frac{\eta\pi\delEq{(-\kappa)^{-1} a}\addZero{r_c}|\ln\varepsilon|\sin^2\theta_e}{\theta_e-\sin\theta_e \cos\theta_e}|\dot{\vec{x}}|^2.
\end{aligned}
\end{equation}
where we have used the \del{facts that $a=\sin\alpha$}\addZero{fact that} and the (normal) velocity of the contact line is given by $\vec{v}_{ct}=\dot{\vec{x}}\cos\theta$.
Notice again that \del{$a$}\addZero{$\alpha$} is a function of $\kappa$ implicitly by~\eqref{generalV0}. The variation of $\Phi$ with respect to $\dot{\vec{x}}$ is given by
\begin{equation}
\begin{aligned}
\label{dPdx}
\frac{ \delta \Phi}{\delta \dot{\vec{x}}} = \frac{2\eta\pi \delEq{(-\kappa)^{-1} a}\addZero{r_c}|\ln\varepsilon|\sin^2\theta_e}{\theta_e-\sin\theta_e \cos\theta_e}\dot{\vec{x}},
\end{aligned}
\end{equation}
\addZero{with $r_c=(-\kappa)^{-1}\sin\alpha \approx (3V_0/\pi Z_1)^{1/3}$. }
The equation corresponds to a friction force of the droplet.
}{\color{black} Notice that \eqref{e:Phi} is computed by integrating a two-dimensional formula along the circular contact line. It is
an approximation to the dissipation in a  three dimensional  mesoscopic annular region near the contact line.
The approximation is good when the width of
the annular region is much smaller  than the radius of the contact line.}

\subsection{The dynamic equation}
When the inertial effect can be ignored, we can use the Onsager principle to derive the dynamic equation for
the location $\vec{x}$  of the droplet on  the substrate $\Gamma$. By the Onsager principle, the motion of the droplet
is determined by minimizing the Rayleighian $\mathcal{R}=\Phi+\dot{\mathcal{E}}$ with respect to the
velocity $\dot{\vec{x}}$ of the droplet on the surface. The corresponding Euler-Lagrange equation is
\begin{equation*}
\frac{\delta\Phi}{\delta\dot{\vec{x}}}+\frac{\delta \mathcal{E}}{\delta \vec{x}}=0.
\end{equation*}
This is a force balance equation between the capillary force $-\frac{\delta \mathcal{E}}{\delta \vec{x}}$ and the viscous friction force $\frac{\delta\Phi}{\delta\dot{\vec{x}}}$.
Combing with the analysis in previous subsections,  i.e. \eqref{e:Edx} and \eqref{dPdx},
we derive a dynamic equation for the location of the droplet on $\Gamma$,
\begin{equation}
\label{gvs}
\dot{\vec{x}}= -\frac{\theta_e-\sin\theta_e \cos\theta_e}{2\eta\pi (-\kappa)^{-1}\delEq{a}\addZero{\sin\alpha}|\ln\varepsilon|\sin^2\theta_e}\frac{d\mathcal{E}}{d \kappa}\nabla_{\Gamma}\kappa.
\end{equation}
It is an ordinary differential equation defined intrinsically  on $\Gamma$, where $\frac{d\mathcal{E}}{d \kappa}$ is defined in \eqref{generaldEdk}.
{\color{black} Here $\nabla_{\Gamma}\kappa$ is the surface gradient of the mean curvature on the surface. It points to a direction in
tangential plane of the substrate where the mean curvature increases.}
One can verify that $\frac{d\mathcal{E}}{d \kappa}$ is always negative, thus the  equation~\eqref{gvs} implies that
the droplet moves in the direction of increasing mean curvature.
{\color{black} That is easy to understand since the total interface energy
in the system will decrease when the droplet moves in this direction.
For example, if the mean curvature of a substrate has different signs,
  a droplet on it tends to move from a  convex region(with a negative mean curvature)
to a concave region(with a positive curvature). Furthermore, the dissipation function contributes an effective
friction coefficient in the equation.
The droplet can move very fast when
the local mean curvature changes dramatically on the substrate. The theory may be helpful to design
surfaces which lead to efficient directional droplet motion in some industrial applications.
}

The equation \eqref{gvs} is a complicated ordinary differential equation since the parameter $\alpha$ is implicitly determined by $\kappa$
through~\eqref{generalV0}.
\addZero{When the droplet is small, by ignoring some higher order terms with respect to $\sin\alpha$,}
 \del{In this case, we can do asymptotic analysis to the equation~\eqref{gvs}(see  details in Appendix~B). The}
 \addZero{the}  equation  is
reduced to (see  details in Appendix~B)
\begin{equation}\label{e:reducedEq}
\dot{\vec{x}}=\frac{\gamma(\theta_e-\sin\theta_e\cos\theta_e)}{4\eta|\ln\varepsilon|\sin\theta_e }r_c^2
\nabla_{\Gamma} \kappa.
\end{equation}
where $r_c=(\frac{3 V_0}{\pi Z_1})^{\frac{1}{3}}$ and  $Z_1=\frac{\sin\theta_e(2+\cos\theta_e)}{4\cos^4(\theta_e/2)}$. 
This is the dynamic equation  for the motion of  a small droplet  driven by the mean curvature
gradient $\nabla_{\Gamma} \kappa({x})$ on a general surface $\Gamma$. It is easy to solve numerically once the parameters, like the Young's angle $\theta_e$, the surface tension $\gamma$, the viscosity $\eta$ and the volume $V_0$, are known.

\add{
\subsection{Models with contact angle hysteresis}
In reality, the equilibrium  contact angle of a liquid on a solid surface may not be unique
due to the geometric roughness or chemical inhomogeneity of the substrate\cite{johnson1964contact,joanny1984model,giacomello2016wetting,marmur2017contact, chatterjee2017surface,XuWang2020}.
The difference between the advancing and the receding contact angle is called contact
angle hysteresis(CAH).
The CAH exists even on a surface with inhomogeneity in nanoscale\cite{giacomello2016wetting,chatterjee2019wetting}.
The CAH induces extra dissipations due to the stick-slip motion of the contact line \cite{joanny1984model}
or the viscous dissipation in the thin films\cite{chatterjee2017surface}.
This will lead to an extra friction force. 

In the following, we consider the contact angle hysteresis effect as in \cite{LvSubstrate}. If the advancing contact angle
$\theta_a$ and the receding contact angle $\theta_r$ are given, the extra friction force can be written as  \cite{extrand1995liquid}
\begin{equation}\label{e:pinningforce}
F_h=k r_c \gamma(\cos\theta_r-\cos\theta_a),
\end{equation}
where $r_c \approx (3 V_0/\pi Z_1)^{1/3}$ and
$k={4}/{\pi}$ is a numerical parameter. Then the dynamic equation~\eqref{gvs} is modified to
\begin{equation}
\label{gvsCAH}
\dot{\vec{x}}= \frac{\theta_e-\sin\theta_e \cos\theta_e}{2\eta\pi (-\kappa)^{-1}\sin\alpha |\ln\varepsilon|\sin^2\theta_e}
\left(-\frac{d\mathcal{E}}{d \kappa}\nabla_{\Gamma}\kappa-F_h \vec{m}\right),
\end{equation}
where $\theta_e=(\theta_a+\theta_r)/2$ and $\vec{m}=\nabla_{\Gamma}\kappa/|\nabla_{\Gamma}\kappa|$. Noticed that the droplet can  move only when the driven force $-\frac{d\mathcal{E}}{d \kappa}|\nabla_{\Gamma}\kappa|$ is larger than $F_h$(we assume the condition always holds in the following). Otherwise, the droplet will be pinned on the surface and $\dot{\vec x}=0$.
Moreover, the reduced model \eqref{e:reducedEq}  becomes
\begin{equation}
\label{e:reduceEqCAH}
\dot{\vec{x}}=\frac{\gamma(\theta_e-\sin\theta_e\cos\theta_e)}{4\eta|\ln\varepsilon|\sin\theta_e }
\left[\Big(\frac{3 V_0}{\pi Z_1}\Big)^{\frac{2}{3}}  \nabla_{\Gamma} \kappa-\frac{2k}{\pi\sin\theta_e}(\cos\theta_r-\cos\theta_a)\vec{m}\right].
\end{equation}

Finally, we would like to remark that some experiments show that the receding contact angle may not be a constant
when there exists retention of
a thin film behind the moving contact line\cite{lam2001effect,lam2002study,chatterjee2017surface}.
In this case,  the interface energy in the receding part may vary in the dynamics of the problem,
since the thickness
of the thin film might depend on the local curvature of the substrate.
This will correspond to a varying receding contact angle in the above model.
We will discuss the effect qualitatively
in next section for the conical surface case.
}
\section{Analysis for the conical surface case}
\subsection{Droplet motion  on a conical surface}
\begin{figure}[H]
  \centering
  \includegraphics[width=12cm]{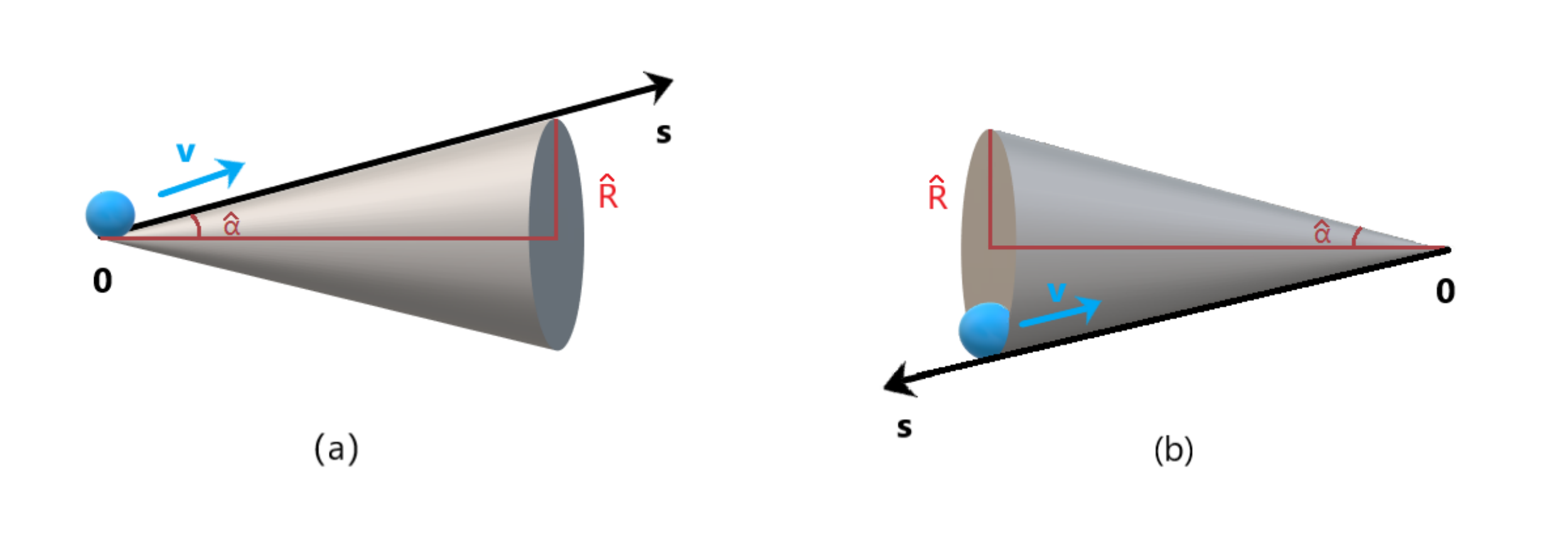}\\
  \caption{(a) and (b) show the droplet motion  on the outside and inside of the cone surfaces, respectively.}\label{01}
\end{figure}

We apply the analysis in Section 2 to the special case of conical surfaces. This problem has been studied experimentally and analytically  in \cite{LvSubstrate,galatola2018spontaneous,JohnDynamics}.
We first consider the droplet motion on the outside surface of a cone with the half apex angle $\hat{\alpha}$(Fig.~\ref{01}).
The  mean curvature on the outside  surface of the cone is given by $\kappa =-(2s\tan\hat{\alpha})^{-1}$, which is
a function of the distance $s$ of a point to the apex.  Direct calculations give that
 $\nabla_{\Gamma}\kappa=(2\tan\hat{\alpha})^{-1} s^{-2} \vec{r}$.
 where $\vec{r}$ is the outward radial direction in the conic surface. By the analysis in the above section,
 we know that the droplet will move to the base on the outside surface. Similar analysis show that
 the droplet moves to the apex on the inside surface of the cone. In the following, we will present some quantitative calculations
only for the outside surface case.

When a droplet of volume $V_0$ is placed on the outside surface of the cone, Eqs.~(\ref{eq:generalU}) and (\ref{generalV0}) reduces into:
\addZero{\begin{align}
\mathcal{E}&=8\pi\gamma s^2\tan^2\hat{\alpha}\left[\frac{\sin^2\alpha}{1+\cos\hat{\theta}}-(1-\cos\alpha)\cos\theta_e\right].\label{Ueqn}\\
V_0&=\frac{8\pi s^3\tan^3\hat{\alpha}}{3}\displaystyle\left[ \sin^3\alpha \frac{(1-\cos\hat{\theta})(2+\cos\hat{\theta})}{\sin\hat{\theta}(1+\cos \hat{\theta})}- (2-3\cos\alpha+\cos^3\alpha)\right]. \label{cseqn}
\end{align}}
They are formulae respectively for the total surface energy and the volume of the droplet when it is located at $s$.
Since $\frac{d \mathcal{E}}{d s}=\frac{d \mathcal{E}}{d \kappa}\frac{d \kappa}{d s}=\frac{1}{2s^2\tan\hat{\alpha}}\frac{d \mathcal{E}}{d \kappa}$, so by Eq.~(\ref{generaldEdk}) we get the derivative of the surface energy with respect to $s$:
\addZero{\begin{equation}
\begin{aligned}
\label{us}
\frac{d \mathcal{E}}{d s}=&\frac{3\gamma V_0}{A s^2\tan\hat{\alpha}} \left[\frac{-2\cos\alpha}{1+\cos\hat{\theta}}-\frac{\sin\alpha\sin\hat{\theta}}{(1+\cos\theta)^2}+\cos\theta_e\right]\\
&+16\pi\gamma s \tan^2\hat{\alpha}\left[\frac{\sin^2\alpha}{1+\cos\hat{\theta}}-(1-\cos\alpha)\cos\theta_e\right].
\end{aligned}
\end{equation}}
where $A$ is defined in Eq.~(\ref{A}). Asymptotic analysis  shows that the generalized force $\frac{d \mathcal{E}}{d s}$ is equivalent to
 that in~\cite{galatola2018spontaneous,JohnDynamics} in leading order when $s$ is large.

Similarly, we can compute the energy dissipation
function for the moving droplet on a conical surface. We insert $\kappa =-(2s\tan\hat{\alpha})^{-1}$ into Eq.~(\ref{e:Phi}) and
noticing  $\dot{\vec{x}}=\dot{s}\vec{r}$, then get:
\begin{equation*}
{\Phi}=\frac{2 \pi\eta \delEq{a}\addZero{\sin\alpha}|\ln\varepsilon|\sin^2\theta_e\tan\hat{\alpha}}{\theta_e-\sin\theta_e \cos\theta_e}s \dot{s}^2.
\end{equation*}
This further leads to
\begin{equation}
\label{phis}
\frac{d \Phi}{d\dot{s}}=\frac{4\pi \eta \delEq{a}\addZero{\sin\alpha}|\ln\varepsilon|\sin^2\theta_e\tan\hat{\alpha}}{\theta_e-\sin\theta_e \cos\theta_e} s\dot{s}.
\end{equation}

When the inertial effect can be ignored, by using the Onsager principle with respect to the velocity $v=\dot{s}$,
 we obtain the moving velocity $v$ of the droplet on the surface,
\begin{equation}
\label{v}
\dot{s}=-\frac{\theta_e-\sin\theta_e \cos\theta_e}{4\pi\eta \delEq{a}\addZero{\sin\alpha}  |\ln\varepsilon|\sin^2\theta_e \tan\hat{\alpha} s}\frac{d \mathcal{E}}{d s},
\end{equation}
where $\frac{d \mathcal{E}}{d s}$ is given in~\eqref{us} and \del{$a$}\addZero{$\alpha$} is  a variable  depending implicitly on $s$
through the equation~\eqref{cseqn}. This is a first order ordinary differential equation for $s$.
 The model is similar to that derived in~\cite{galatola2018spontaneous}.
The main difference is that we take into account the viscous dissipation of the droplet to give an exact formula for the (effective) friction coefficient,
while it is a fitting parameter in~\cite{galatola2018spontaneous}.

When the inertia is not negligible, the capillary forces $-\frac{d \mathcal{E}}{d s}$ and the viscous friction forces $\frac{d \Phi}{d\dot{s}}$
are not balanced.
As in \cite{LvSubstrate} and \cite{galatola2018spontaneous}, we can
 assume that the acceleration of the droplet motion is driven by the combination of the capillary force and the viscous friction force. This leads to
\begin{equation}\label{e:forcebalance}
  m\frac{dv}{dt}+\frac{d \Phi}{d \dot{s}}+\frac{d \mathcal{E}}{d s}=0,
\end{equation}
or equivalently,
\begin{equation}\label{e:ODE2}
\frac{d v}{dt }+\frac{4\pi\eta \delEq{a}\addZero{\sin\alpha} |\ln\varepsilon|\sin^2\theta_e\tan \hat{\alpha}s}{\rho V_0 (\theta_e-\sin\theta_e \cos\theta_e)}v+\frac{1}{\rho V_0}\frac{d \mathcal{E}}{d s}=0,
\end{equation}
where $\rho$ is the density of the liquid and $v=\dot{s}$. This is actually  a second order differential equation for $s$, which
 can be solved once the initial location and the initial velocity of the droplet are given.
 To compare with the experimental data in \cite{LvSubstrate}, we  change the form of the equation~\eqref{e:ODE2}.
Using the relations $\hat{R}=s\sin\hat{\alpha}$ and $d\hat{R}/dt =\sin\hat{\alpha} ds/dt=v \sin\hat{\alpha} $(see Fig.~\ref{01}),
the equation~\eqref{e:ODE2} can be rewritten as,
\begin{equation}
\label{dRdv}
  \frac{dv}{d\hat{R}}+\frac{8\pi\eta  \delEq{a}\addZero{\sin\alpha} |\ln\varepsilon|\sin^2\theta_e\hat{R}}{\rho V_0 \sin (2\hat{\alpha})(\theta_e-\sin\theta_e \cos\theta_e)}+\frac{1}{\rho V_0 v}\frac{d \mathcal{E}}{d \hat{R}}=0,
\end{equation}
where we have used the fact $\frac{d \mathcal{E}}{d s}=\frac{d \mathcal{E}}{d \hat{R}} \sin\hat{\alpha}$.

\add{
When there exists contact angle hysteresis, the equation~\eqref{e:forcebalance} becomes
\begin{equation}
  m\frac{dv}{dt}+\frac{d \Phi}{d \dot{s}}+F_h+\frac{d \mathcal{E}}{d s}=0,
\end{equation}
and it is equivalent to
\begin{equation}\label{e:final}
  \frac{dv}{d\hat{R}}+\frac{8\pi\eta  \sin\alpha|\ln\varepsilon|\sin^2\theta_e\hat{R}}{\rho V_0 \sin (2\hat{\alpha})(\theta_e-\sin\theta_e \cos\theta_e)}+\frac{F_h}{\rho V_0 v \sin{\hat{\alpha}}}+\frac{1}{\rho V_0 v}\frac{d \mathcal{E}}{d \hat{R}}=0,
\end{equation}
where $F_h$ is given in~\eqref{e:pinningforce} and $\theta_e=(\theta_a+\theta_r)/2$.
The difference between the model~\eqref{e:final} from that in \cite{LvSubstrate} is that the viscous dissipations
in the droplet are considered here while it is ignored in the previous work.
The solution of the equation exists a maximal velocity at some critical
radius $R_{c}$ even when $F_{h}=0$.
Actually, Eq.~\eqref{v} show that the velocity is a monotone decreasing function with respect to $\hat{R}$ if we do not consider the inertial effect.
When we further consider the dynamics of the droplet from a static state,  Eq.~\eqref{dRdv}  implies that
the velocity first increases due to the acceleration and then decreases when the driven force is balanced by the viscous friction forces.
}

\subsection{Long time behavior of a sliding droplet on a conical surface}
In this subsection, we  
show that the motion of the droplet on the outside surface of a cone satisfies a power law
that $s\sim t^{1/3}$ for relatively late time $t$.
When $s$ is  large enough, the mean curvature $\kappa =-(2s\tan\hat{\alpha})^{-1}$ becomes so small
 that $V_0|\kappa|^3\ll 1$. Then the reduced model~\eqref{e:reducedEq} 
 applies.
Noticing again that $\vec{x}=s \vec{r}$ and $\nabla_{\Gamma}\kappa=(2\tan\hat{\alpha})^{-1} s^{-2} \vec{r}$,
the equation~\eqref{e:reducedEq} is reduced to
\begin{equation*}
  \dot{s}=\frac{\gamma (\theta_e-\sin\theta_e \cos\theta_e)}{8 \eta |\ln\varepsilon|\sin\theta_e  \tan \hat{\alpha}}\Big(\frac{3 V_0}{\pi Z_1}\Big)^{\frac{2}{3}} s^{-2},
\end{equation*}
where $Z_1=\frac{\sin\theta_e(2+\cos\theta_e)}{(1+\cos\theta_e)^2}$.
This leads to
$$3s^2\dot{s}=\sigma^3,$$
 where
$\sigma^3=\frac{3\gamma (\theta_e-\sin\theta_e \cos\theta_e)(3V_0)^{2/3}}{8 \pi^{2/3}\eta |\ln\varepsilon|\sin\theta_e Z_1^{2/3} \tan\hat{\alpha}}$.
Integration of the equation gives
\begin{equation}
\label{sstt}
s^3=s_0^3+\sigma^3 t,
\end{equation}
where $s_0$ is the initial position of the droplet center. It shows that $s \sim t^{1/3}$ for sufficiently late time.

In  Ref.~\cite{JohnDynamics}, McCarthy {\it et al.} study the long time dynamics of a water droplet on a conical surface
covered with a silicone oil film.  They reported a slower power law that $s\sim t^{1/4}$, since the
energy dissipation in the oil films dominate in their problem. Our analysis shows that the long time behaviour of
a droplet on a clean conical surface, the position of the droplet is characterized by
a scaling law $s\sim\sigma t^{1/3}$. {\color{black} The results can be understood as follows.
 When a droplet moves
on a substrate coated with thin oil films,  the motion may induce flows in the oil film.
This leads to extra energy dissipations in the system and  a slower motion of the droplet.}
\add{Detailed theoretical analysis for the problem with an oil covered substrate is presented in \cite{JohnDynamics}.}

\add{
\subsection{Discussions on the thin film effect}
As mentioned in Section 2, thin film retention may occur in some systems behind the moving contact
line due to the long-ranged intermolecular interaction between the solid and the liquid.
The thin film change the free energy in the receding part\cite{lam2001effect,lam2002study,chatterjee2019wetting}.
In this subsection, we will discuss briefly how the local curvature of the substrate  affects
the thickness of the thin film and the receding contact angle.

It is known that the effective surface energy density $\gamma^{film}_{SV}$ of a thin film covered solid surface
is given by~\cite{Gennes03,chatterjee2019wetting}
$$\gamma_{SV}^{film}=\gamma_{SL}+\gamma+e(h),$$
where $e(h)$ is the excess energy of the thin film. $e(h)$ is a function of the film thickness $h$.
Notice that the receding contact angle
$\theta_r$
is given by
$
\gamma_{SV}^{film}-\gamma_{SL}=\gamma\cos\theta_r.
$
Then the receding contact  angle can be computed as
\begin{equation}\label{e:recedingCA0}
\cos\theta_{r}=1+\frac{e(h)}{\changeOne{\gamma}}.
\end{equation}
The standard theory shows that the excess energy $e(h)$ is given by \cite{Gennes03,chatterjee2019wetting}
\begin{equation}\label{e:excess}
e(h)=\frac{\mathrm{A}_{L-SL}}{12\pi h^2},
\end{equation}
where $\mathrm A_{L-SL}$ is the Hamaker constant of interaction between the liquid surface  and solid-liquid surface \cite{chatterjee2019wetting}.
\changeOne{The constant can be computed  by  $\mathrm A_{L-SL} \approx(\sqrt{A_{S}} - \sqrt{\mathrm A_{L}})(\sqrt{\mathrm A_{V}} - \sqrt{\mathrm A_{L}})$, where $\mathrm A_{S}$, $\mathrm A_{L}$ and $\mathrm A_{V}$ are corresponding Hamaker constants in the system~\cite{1992Intermolecular}. For a water surface on a glass substrate, one has $\mathrm A_{S} \approx 6.8 \times 10^{-20}$J, $\mathrm A_{L} \approx 3.7 \times 10^{-20}$J and $\mathrm A_{V} \approx 0$J \cite{2020How}.}

To determine $\theta_r$, we need only to compute the thickness of the thin film.
For that purpose, we consider  the \changeOne{disjoining pressure}
\changeOne{
\begin{equation}\Pi(h)=\frac{\partial e(h)}{\partial h}=-\frac{\mathrm A_{L-SL}}{6\pi h^3}.
\end{equation}
}
As in~\cite{keiser2018dynamics},  the \changeOne{disjoining pressure $\Pi(h)$} is equal to the pressure \changeOne{$p$} inside the droplet in equilibrium state. By the Young-Laplace equation, we have
\changeOne{
\begin{equation}
p=\frac{2\gamma}{r},
\end{equation}
}
where $r$ is the radius of the droplet. This  leads to $\changeOne{h=(\frac{-r\mathrm A_{L-SL} }{12\pi\gamma})^{1/3}}$.  By Eqs.~\eqref{e:recedingCA0} and \eqref{e:excess},
we have
\begin{equation}\label{e:recedingCA1}
\cos\theta_r=1+\left(\frac{\mathrm A_{L-SL}}{12\pi \gamma}\right)^{\frac{1}{3}} r^{-\frac{2}{3}}. 
\end{equation}
When a droplet moves outwards on the out surface of a cone, the radius of the droplet will increase gradually.
By the above equation, we know that the receding contact angle will decrease since the Hamaker constant $\mathrm A_{L-SL}$ is negative.
Then the friction force $F_h$  becomes larger and
 this tends to slow down the motion of the droplet.
However, when the droplet is very small with respect to the  curvature radius of the droplet (i.e. $V_0|\kappa|^3 \ll 1$),
then $r\approx r_c/\sin\theta_e$ in leading order and \eqref{e:recedingCA1} implies that $\theta_r$
can be approximated well by a constant value.
}
\section{Numerical results}
\subsection{Comparison with the experimental data}\label{Comparison}
To verify the analytical results in previous sections, we first do some comparisons with the experimental results in~\cite{LvSubstrate}.
We consider a small water droplet moving on the outside surface of a \changeOne{glass} cone.
The physical parameters of water are selected at $20\degree \mathrm{C}$, which are the same as the experiment:
 the viscosity $\eta = 1.0087\times10^{-3}\mathrm{Pa}\cdot \mathrm{s}$,
the density $\rho = 0.998232\times10^3\mathrm{kg}/\mathrm{m}^3$, the surface tension $\gamma = 7.280\times10^{-2}\mathrm{N}/\mathrm{m}$,
and the gravitational acceleration parameter $g=9.8\mathrm{m}/\mathrm{s}^2$.
The capillary length is calculated as $L_c=\sqrt{\gamma/\rho g}=0.0027\mathrm{m}$. Here we simply set $\ln\varepsilon=O(10)$(see \cite{Gennes03}), \add{and $\theta_e=(\theta_a+\theta_r)/2$}.
\begin{figure}[H]
  \centering
  \includegraphics[width=9cm]{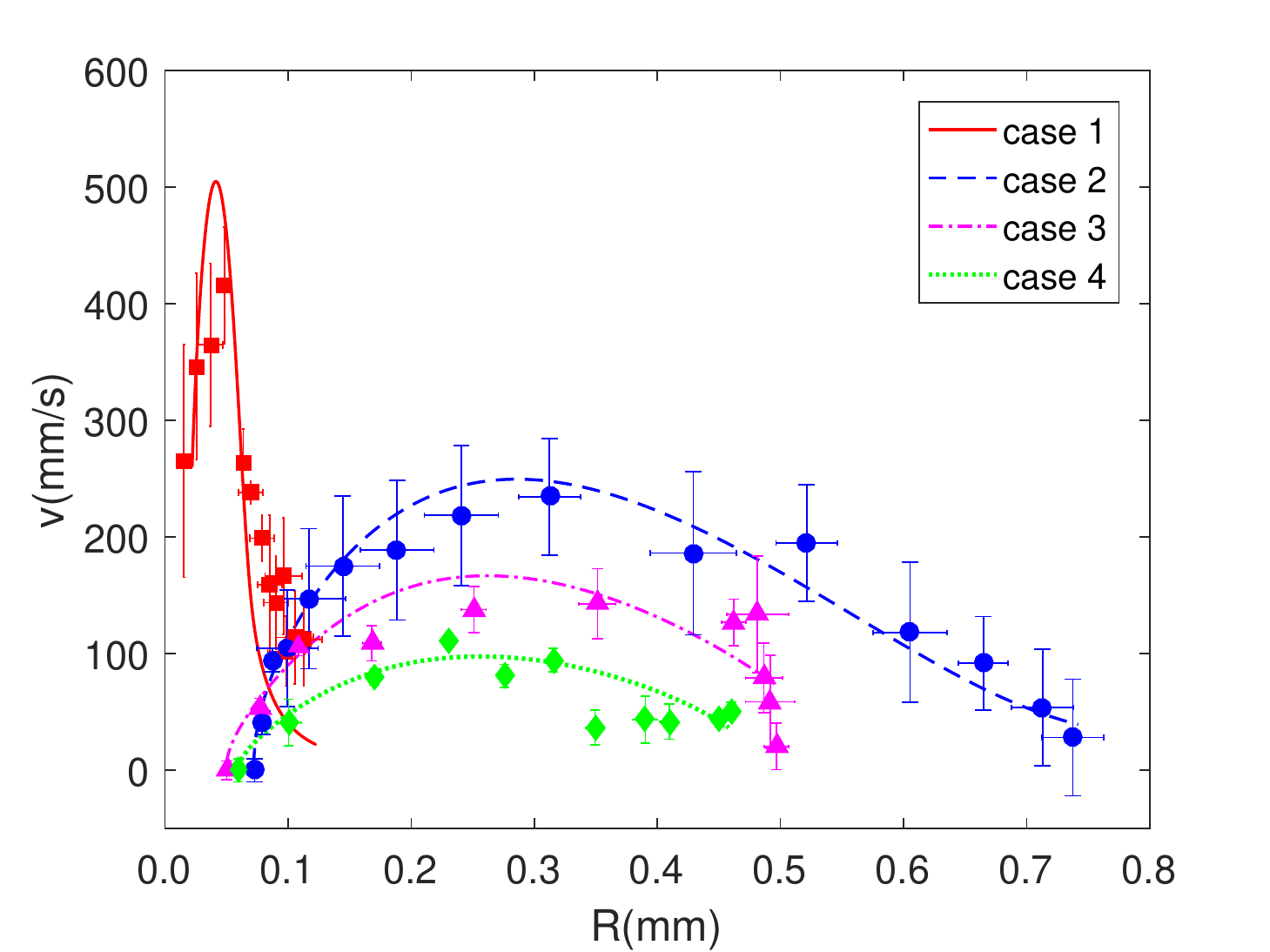}\\
  \caption{\add{Comparison between the experimental and the analytical results for speeds of the droplets as a function of the local radius $\hat{R}$ of the circular section of the cone.
The points with error bars corresponds to experiments deduced from Fig. 1(b) of Ref. \cite{LvSubstrate} and the lines are computed from the reduced models of Eqs. \eqref{e:final}. From top to bottom, squares and red full line: $V_{0}=1.15 \times$ $10^{-2} \mathrm{mm}^{3}, \hat{\alpha}=5.1^{\circ}, \theta_a=10.8^{\circ},
\theta_r=9.3^{\circ}, v_0=260\mathrm{mm/s}, \hat{R}_{\min }=2.2 \times 10^{-2} \mathrm{mm}, \ln\varepsilon=15$; circles and blue short-dashed line: $V_{0}=1 \mathrm{mm}^{3}, \hat{\alpha}=3.2^{\circ}, \theta_a=55.3^{\circ}, \theta_r=53.6^{\circ}, v_0 =3.5\times10^{-3} \mathrm{mm/s},  \hat{R}_{\min }=7.21\times10^{-2}\mathrm{mm}, \ln\varepsilon=10$; triangles and purple long-dashed line: $V_{0}=1 \mathrm{mm}^{3}, \hat{\alpha}=3.2^{\circ}, \theta_a=69.5^{\circ}, \theta_r=65^{\circ}, v_0 =6.7\times10^{-3} \mathrm{mm/s}, \hat{R}_{\min }=5 \times 10^{-2} \mathrm{mm}, \ln\varepsilon=10$; and diamonds and green dot-dashed line: $V_{0}=1 \mathrm{mm}^{3}, \hat{\alpha}=3.2^{\circ}, \theta_a=88^{\circ}, \theta_r=85^{\circ}, v_0 =2.5\times10^{-3} \mathrm{mm/s}, \hat{R}_{\min }=6 \times 10^{-2} \mathrm{mm}, \ln\varepsilon=10$.
}}\label{OK}
\end{figure}

When a droplet is placed on the outside of the cone, it moves spontaneously from tip to the base.
We compare the velocity of the droplet with the experimental data in~\cite{LvSubstrate}.
The results are shown in Figure~\ref{OK}. We consider four cases as in the experiments.
In the first case where the volume of the droplet is
small($V_{0}=1.15 \times 10^{-2} \mathrm{mm}^{3}$), the theoretical prediction fits
the experimental data very well when we choose the advancing and receding contact angles
the same as in the experiments. This is much better than the results shown in \cite{LvSubstrate,galatola2018spontaneous}.
In other cases, the volume of the droplet is relatively large($V_{0}=1 \mathrm{mm}^{3}$),
the numerical results by the reduced model  fit well with the experiments only when we
choose relatively larger contact angles than those in experiments.
All other the parameters are chosen the same as in experiments except the
contact angle $\theta_e$.
The numerical results are similar to that obtained by Galatola in~\cite{galatola2018spontaneous},
where an extra fitting parameter $\xi$ must be chosen  differently in these cases
 to compare with the experiments.

\begin{figure}[H]
  \centering
  \includegraphics[width=9cm]{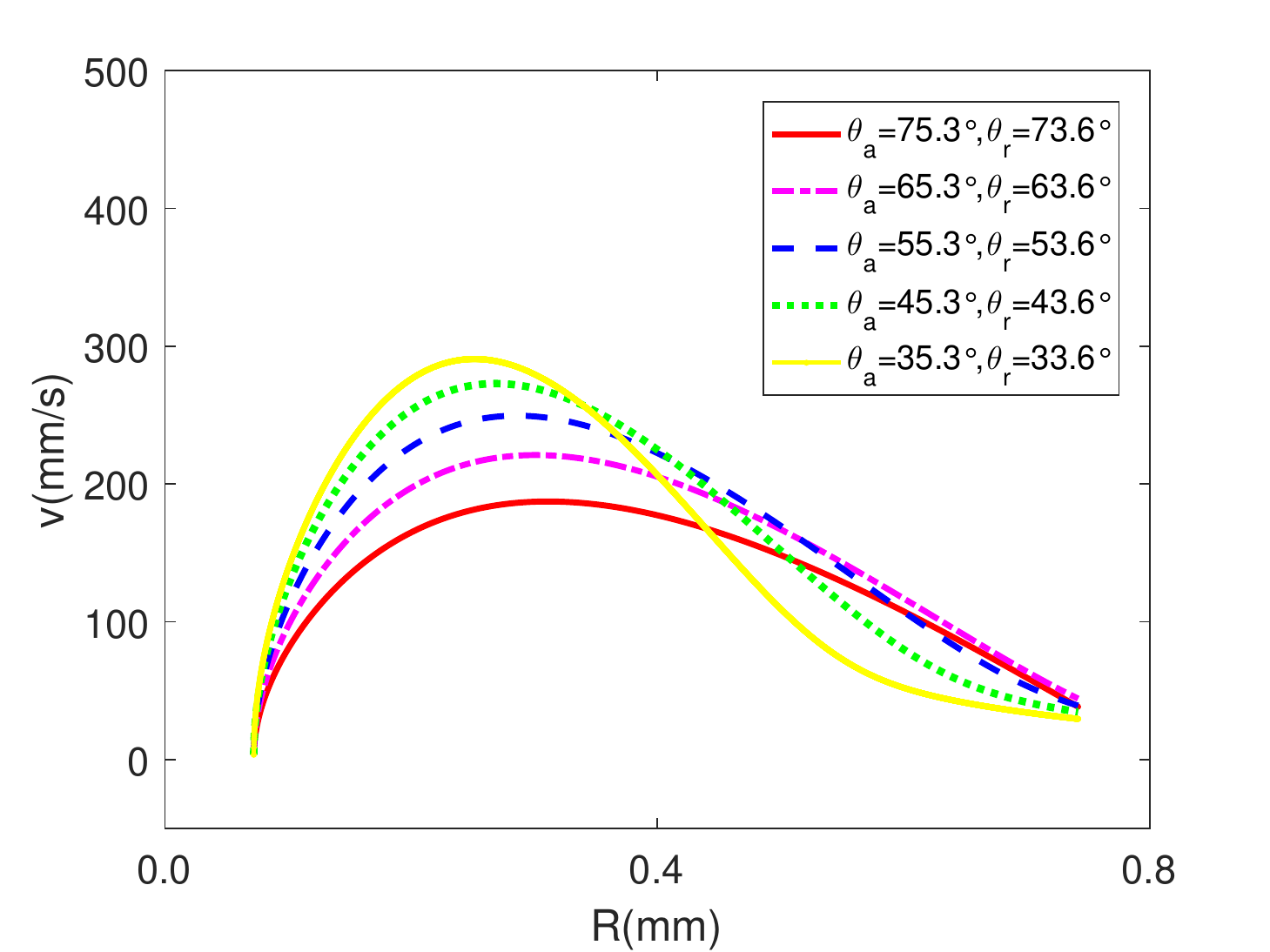}\\
  \caption{\add{The dynamics of the droplet for different contact angles. The other parameters are the same as in Case 2 in Fig. \ref{OK}.}}\label{Test}
\end{figure}

\add{The choice of relatively larger contact angles for  large droplet cases can be understood in following ways.
Firstly, as discussed in~\cite{LvSubstrate,galatola2018spontaneous},
the apparent contact angle is different from the contact angles in equilibrium.
Actually, this can be seen from the experiments in \cite{LvSubstrate}.
In Figure~\ref{Test}, we show the velocity of the droplet for different choice of the contact angles. It is found that the
contact angles can affect the dynamics largely. This indicates that we need to use the  apparent angles in our model if
they are different from the static ones.
Secondly, the  disagreement might also be induced by the large error of the approximation for the total surface energy.
Notice that the cone is very sharp in the experiments. The approximations for the free energy in section 2 might
 not be accurate enough when the volume of the droplet is large.

Finally, we would like to remark that the contact angle hysteresis is very important  to compare with the experiments.
Without considering the contact angle hysteresis force $F_h$, it is much more difficult to use the equation~\eqref{dRdv} to fit with
 the experimental data
unless one choose very large contact angles as in \cite{galatola2018spontaneous}.
}

\begin{figure}[H]
  \centering
  \includegraphics[width=8cm]{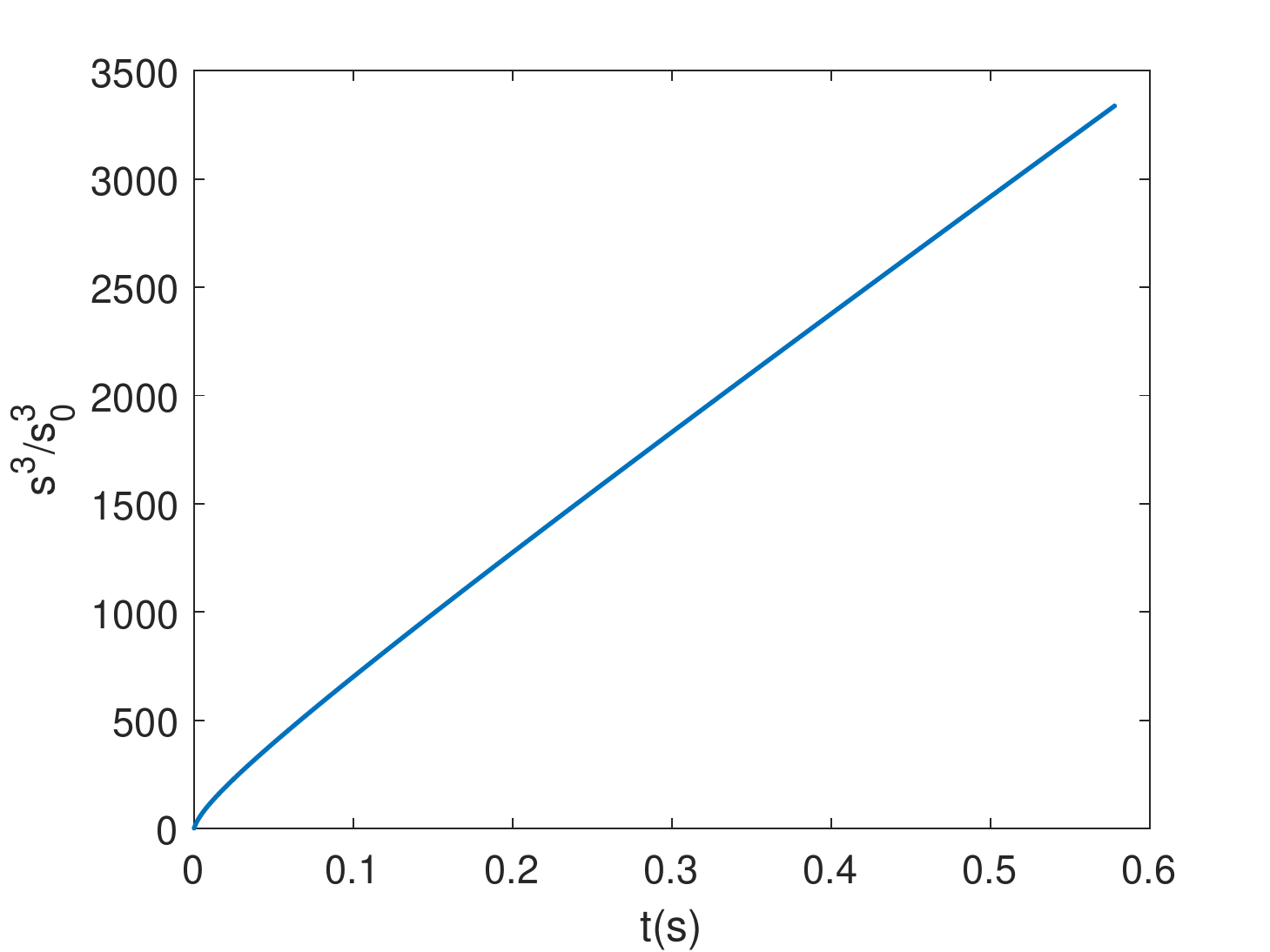}\\
  \caption{A linear relation between $s^3/s^3_0$ and $t$(far from the initial time) verifies the prediction of Eq.~(\ref{sstt}). Here we solve the equation~\eqref{v} by setting $V_0=1.15 \times$ $10^{-2} \mathrm{mm}^{3}, \hat{\alpha}=5.1^{\circ}, \theta=10.05^{\circ}, s_0 = 0.247\mathrm{mm}$. }  \label{st}
\end{figure}

\add{\subsection{The scaling law of the droplet motion on a conical surface}\label{Powerlaw}}
By asymptotic analysis, we have shown a scaling law that $s \sim t^{1/3}$ for the position of the droplet
 in Eq.~(\ref{sstt}) when $t$ is large.
This scaling law is verified numerically as  in Fig.~\ref{st}, where a typical solution of Eq.~\eqref{v} is illustrated.
We could see that $s^3/s_0^3$ is linear to $t$ in the later time. \add{
Experimental verification of the power law is still needed in the future.
}

\begin{figure}[H]
\subfigure[]{
\begin{minipage}[t]{0.5\linewidth}
\centering
\includegraphics[width=2.99in]{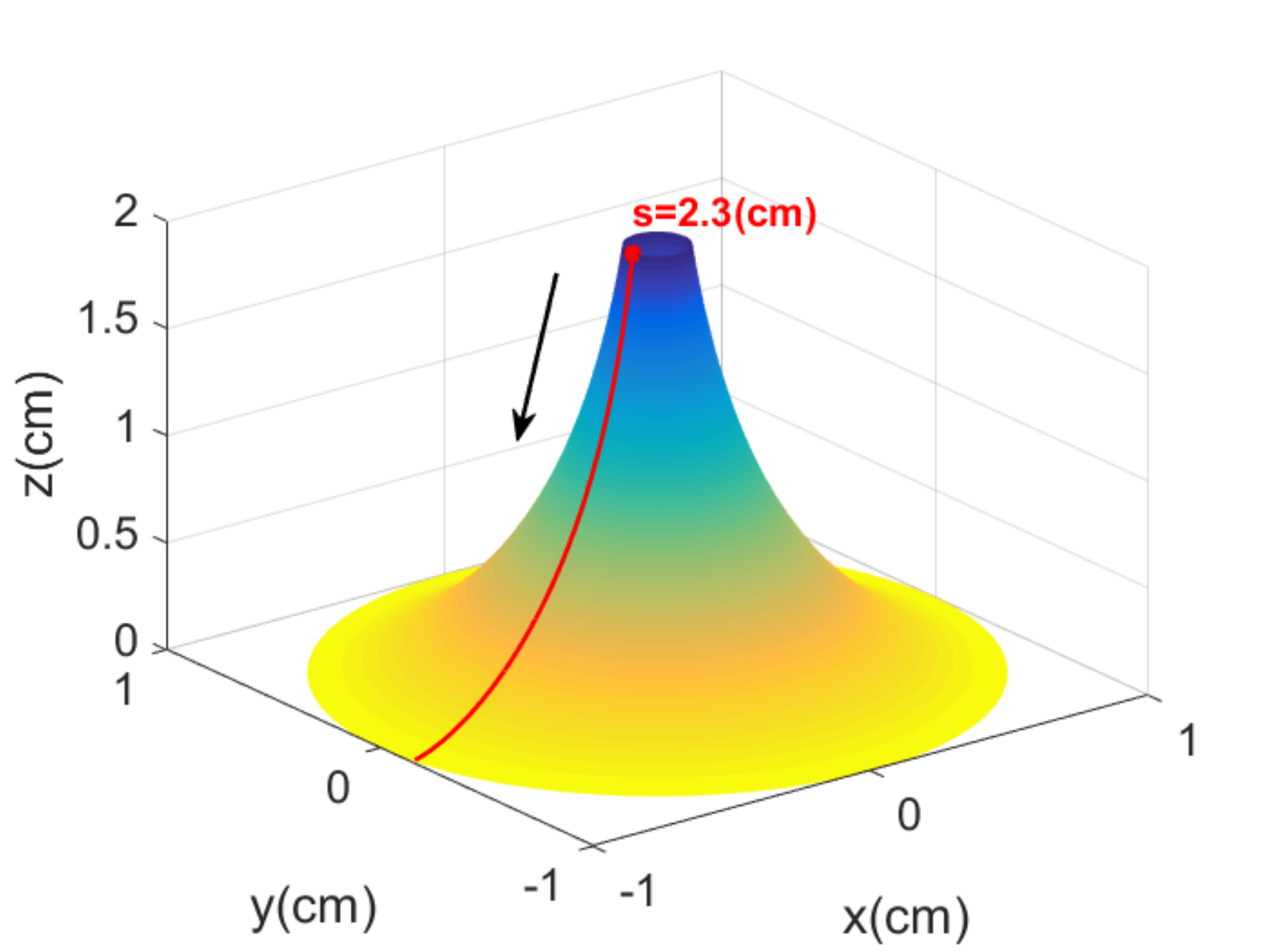}
\end{minipage}
}
\subfigure[]{
\begin{minipage}[t]{0.5\linewidth}
\centering
\includegraphics[width=2.99in]{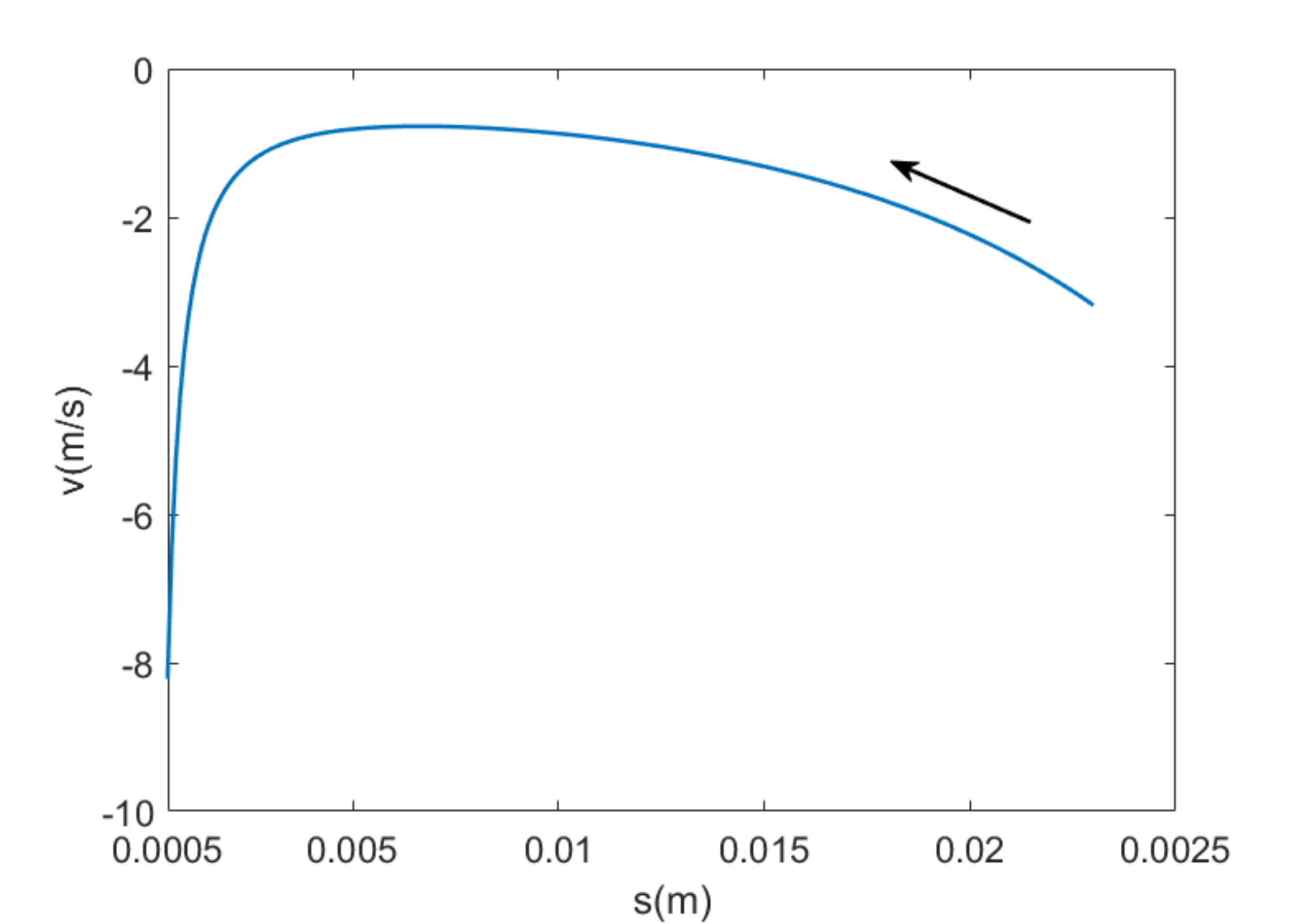}
\end{minipage}%
}
\caption{(a). The trajectory of the a sliding droplet(downwards). (b). The relation between the velocity $v$ and the  arc length $s$ computed by Eq. (\ref{gvs}). Here, we set $b=c=0.01, V_0=1.0 \mathrm{mm}^{3}, \theta=130^{\circ}, s_0 = 2.3\mathrm{cm}$.
We stop the calculation  when $s=0.5\mathrm{mm}$. }\label{test}
\label{F}
\end{figure}

{
\subsection{Droplet motion on general surfaces} 
In this subsection, we will  give some numerical examples for droplet motion on non-conical surfaces.
The  examples indicate that the reduced model proposed in Section 2 can be used to study the droplet motion   on
general smooth substrates.

The first example is an radial symmetric surface as shown in Fig.~\ref{F}.
In parametric coordinates,  the surface is given by
$$\Gamma=\{ \vec{r}(s,\varphi)=(f(s)\cos\varphi,f(s)\sin\varphi,g(s)),\; f(s)>0\}.$$
It is generated by rotating a curve, given by  $\{(f(s),0,g(s))\}$ in arc length parameter $s$, around the $z-$axis.
$\varphi$ is the rotating angle.
Here we set $f(s)=ce^{-s/b}$, $g(s)=\int_{s_0}^s\sqrt{1-\frac{c^2}{b^2}e^{-2\bar{s}/b}}d\bar{s}$, $s\geq 0$.
It is easy to see that the curve intersects with the plane $z=0$ when $s=0$ and extends upward when $s>0$.
Then the mean curvature of the curved surface is a function of $s$ and can be computed as
$$\kappa=\frac{1}{2}\left[(\frac{c}{b^2e^{s/b}}-\frac{e^{s/b}}{c})(1-\frac{c^2}{b^2 e^{2s/b}})^{\frac{1}{2}}+\frac{c^3}{b^4e^{3s/b}}(1-\frac{c^2}{b^2 e^{2s/b}})^{-\frac{1}{2}}\right].$$ Accordingly the derivative is
\begin{align*}\partial_s \kappa
=-\frac{1}{2}&\left[(\frac{e^{s/b}}{bc}+\frac{c}{b^3e^{s/b}})(1-\frac{c^2}{b^2 e^{2s/b}})^{\frac{1}{2}}\right. \\
&+\left.(\frac{c}{b^3 e^{s/b}}+\frac{2c^3}{b^5 e^{3s/b}} )(1-\frac{c^2}{b^2 e^{2s/b}})^{-\frac{1}{2}}+\frac{c^5}{b^7e^{-5s/b}}(1-\frac{c^2}{b^2 e^{2s/b}})^{-\frac{3}{2}}\right]\leq 0.
\end{align*}

 By the analysis in Section 2, if we put a droplet on such a surface, it will moves in the direction of $s$ decreases along a generating
  curve. This means the droplet moves from top to base. In Figure~\ref{F}(a), we draw one trajectory of such a motion.
%
Moreover, according to Eq.\eqref{gvs}, we can compute the relation between the velocity $v$ and arc length $s$, which is shown in Figure~\ref{F}(b).
In comparison with that in Figure~\ref{OK}(for the case $V_0=1\mathrm{mm}^3$), we could see that the sliding velocity is much larger than that on a conical surface, due to the large surface gradient of the mean curvature.
In addition,  the varying behaviour of the velocity is also different due to the effect of the geometry of the substrate.
}
Finally, we  consider a monkey saddle surface given by $\Gamma:=\{\vec{r}\,(u,v) = (u,v,u^3-3uv^2)^T\}$,
where $(u,v)$ are parametric variables.
In this case, Eq.~(\ref{gvs}) can be rewrite as:
\begin{align*}
  \dot{\vec{x}}&= -\frac{\theta_e-\sin\theta_e \cos\theta_e}{2\eta\pi (-\kappa)^{-1}|\ln\varepsilon|\sin^2\theta_ea}\frac{d\mathcal{E}}{d \kappa}\nabla_{\Gamma}\kappa\\
               &= -\frac{\theta_e-\sin\theta_e \cos\theta_e}{2\eta\pi (-\kappa)^{-1}|\ln\varepsilon|\sin^2\theta_ea}\frac{d\mathcal{E}}{d \kappa}D\,\vec{r}\,G^{\,-1}\nabla\kappa,
\end{align*}
where $D\,\vec{r}\,=(\vec{r}_u,\vec{r}_v)$, $G =(D\,\vec{r})^{T}D\,\vec{r}$ is the metric tensor, and $\nabla \kappa=(\frac{\partial \kappa}{\partial u},\frac{\partial \kappa}{\partial v})^{T}.$ The equation can be solved numerically.

\begin{figure}[H]
  \centering
  \includegraphics[width=9cm]{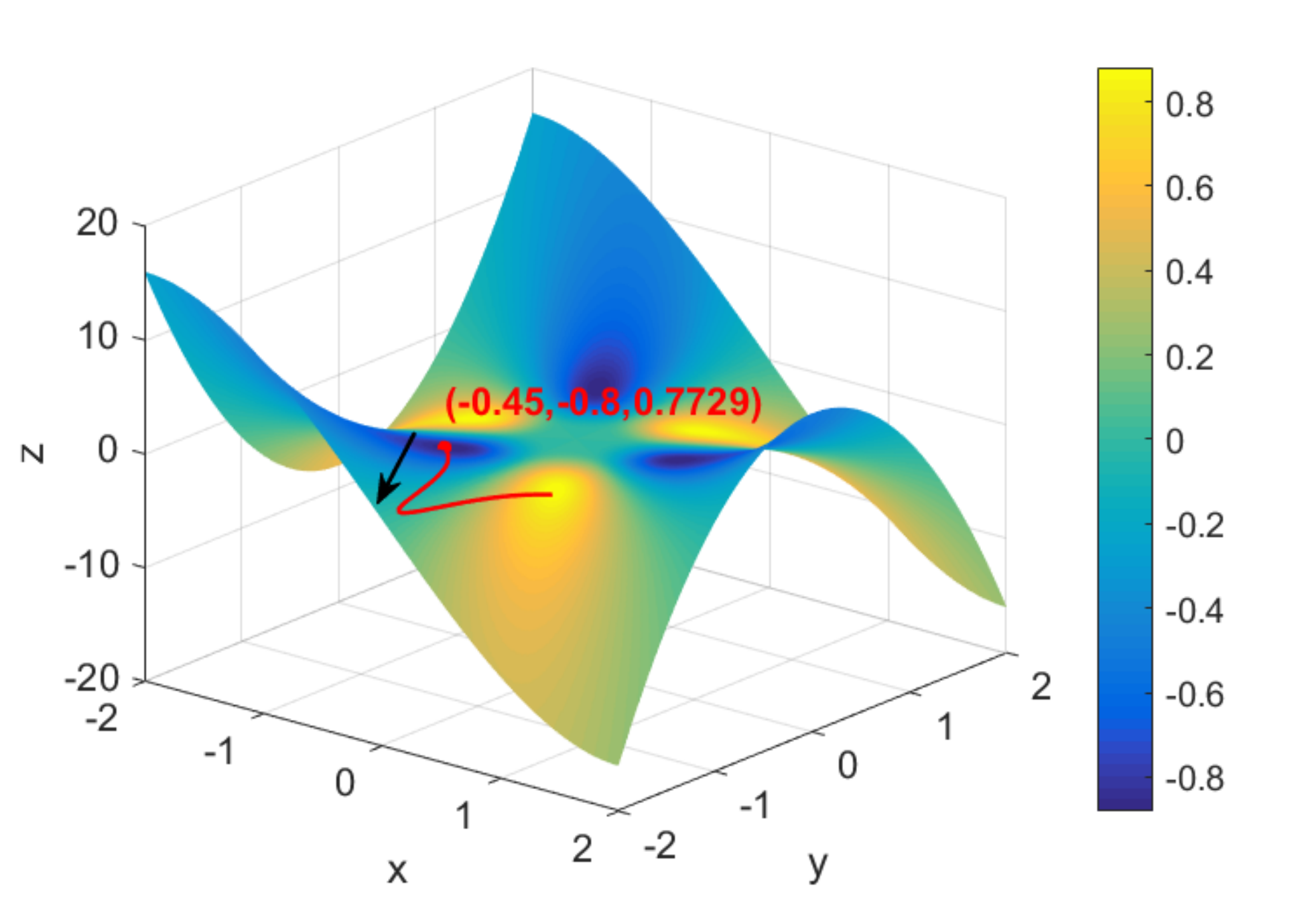}\\
  \caption{A trajectory of a droplet on a monkey saddle surface, the colorbar represents the mean curvaure $\kappa$.}
  \label{Monkey}
\end{figure}

In~Fig.~\ref{Monkey}, we set  a small droplet 
initially on $\vec{r}_0=(-0.45,-0.8,0.7729)$, which is close to a point with locally smallest
mean curvature on the surface. By solving the above equation, we  obtain the  trajectory of the droplet on the surface.
We could see that the droplet moves in a complicated way to reach a point where the surface has largest mean curvature.
Actually, the droplet moves along the surface gradient direction of the mean curvature at each point on the trajectory.
 In this final state, the mean curvature achieves a local maximum(its surface gradient is zero),
 where the system has a locally minimal energy.

\section{Conclusion}
In this paper, we use the Onsager principle as an approximation tool
 to study the spontaneous motion of a small droplet on general curved surfaces.
The droplet motion is driven by the capillary force induced by curvature gradient of the substrate.
We compute approximately the capillary energy and the viscous dissipation in the droplet.
An ordinary differential equation is derived for the
intrinsic displacement of a droplet on a surface.
The equation describes quantitatively how a droplet moves  in the increasing
direction of the mean curvature(along its surface gradient direction).
We present some numerical examples for the trajectories of  small droplets on some complicated surfaces.
We expect that the reduced model may be used to design surfaces with good transport properties.

The reduced model applies directly  to the droplet motion on conical surfaces, which has been studied a lot recently.
It recovers formally the previous model in \cite{galatola2018spontaneous}. The capillary force
is consistent with that in~\cite{galatola2018spontaneous,JohnDynamics} in leading order. However, the effective friction
coefficient in our model is derived from the viscous dissipation in the droplet, not just that of the contact line.
Numerical results by the reduced model are in good agreement with  experiments in Lv {\it et al.} \cite{LvSubstrate}
without adjusting the friction coefficient.
In addition, we derive a  scaling law $s\sim t^{1/3}$ (for sufficient late time) of the displacement of the droplet
 by asymptotic analysis on the conical
surface. The power law is fast than the relation $s\sim t^{1/4}$ reported in \cite{JohnDynamics} since the
dissipation laws are different. In~\cite{JohnDynamics},
the substrate is covered by silicone oil where the dissipation is much larger than that in the droplet.

\changeOne{Some issues in the problem  are not completely studied in our analysis
and need to be further investigated in the future. Firstly, we  consider only
the case when  the droplet is relatively small with respect to the  curvature radius of the substrate.
In this case, the surface energy can be approximated well by that of a droplet on a spherical substrate.
When the droplet is  large, its shape  is not so  simple and
we need to consider the shape changes and use more parameters to
characterize the problem\cite{xu2016variational,mayo2015simulating}.
Secondly, the thin film effect in the receding part is discussed only qualitatively
in this paper. The effect might be important in some situations \cite{lam2001effect,lam2002study,chatterjee2017surface}
and this needs to be studied quantitatively.
In addition, the theoretical prediction on the power law of the motion of a droplet on
conical surfaces also needs to be verified experimentally.
 }

Finally, we would like to remark that the Onsager variational principle is a powerful approximation tool for
theoretical analysis. The analysis can be generated simply to the case for substrates with both geometric and chemical inhomogeneity.
The method  can also be used to deal with more complicated problems,
e.g. dynamic wetting on soft substrates or other geometries (\cite{dervaux2020nonlinear,lv2021wetting}).
These problems will be left for future work.

%
%
%
%
\section*{Acknowledgements}
The work was partially supported by NSFC 11971469 and by the National Key R\&D Program of China under Grant 2018YFB0704304 and Grant 2018YFB0704300.
The authors would like to thank the 
referees for their careful reading of the manuscript of the presented paper and very
valuable comments which help to improve the paper a lot.
\section*{Appendix}
\subsection*{A. Computations for the viscous dissipation in a two dimensional region}
\begin{figure}[H]
  \centering
  \includegraphics[width=2.5in]{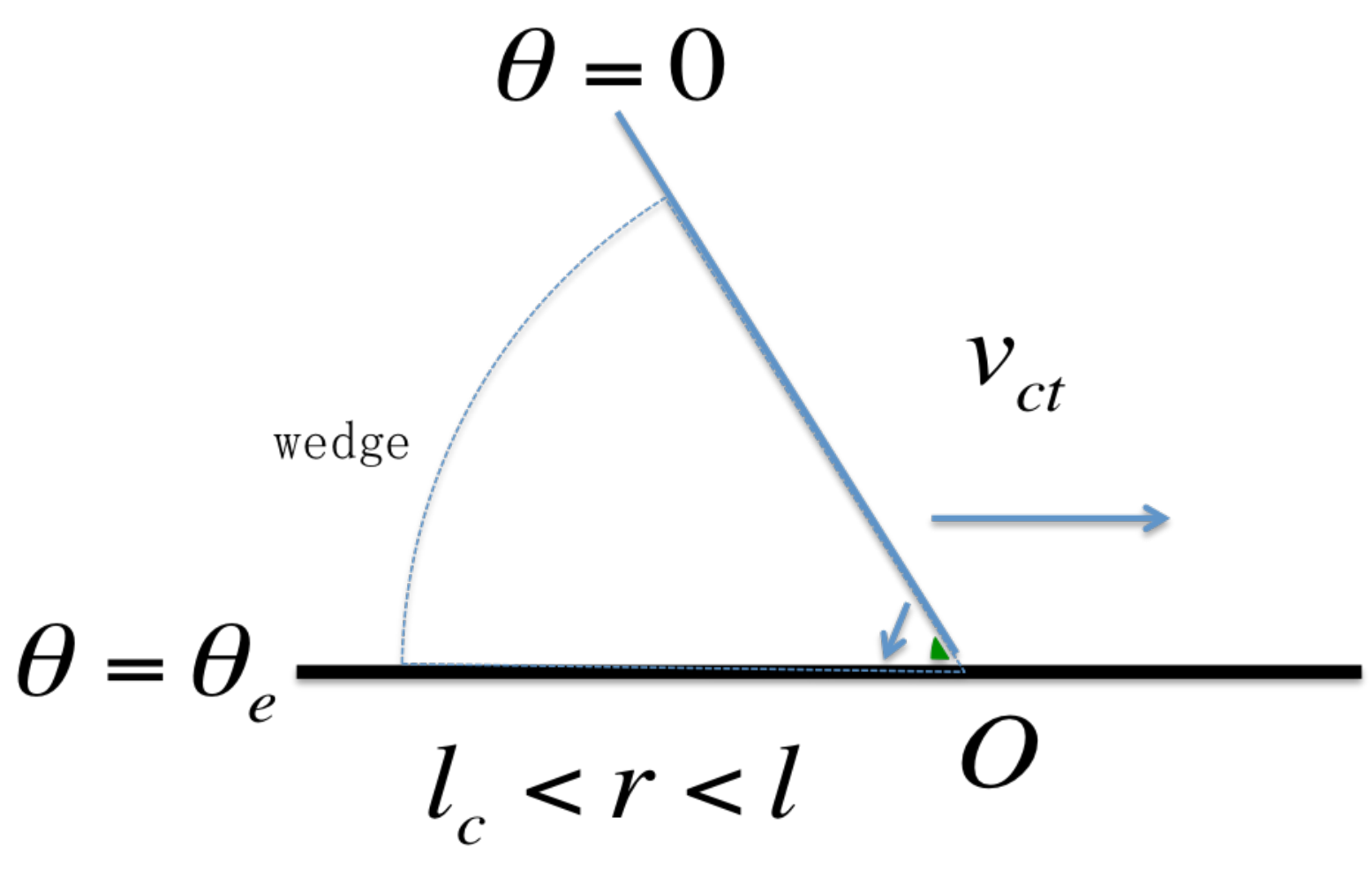}\\
  \caption{A wedge region near a contact point}\label{fig:wedge0}
\end{figure}

\setcounter{equation}{0}
\renewcommand{\theequation}{A\arabic{equation}}
We first study the motion of viscous fluid in a wedge region as shown in Fig.~\ref{fig:wedge0}.
The fluid velocity on the bottom surface is zero by a noslip boundary condition.
The straight line  of the liquid surface moves in the right direction with a velocity $v_{ct}$.
We choose a frame moving with the contact point. Then we can consider the following  Stokes equation,
\begin{equation}\label{e:stokes}
\left\{
\begin{array}{ll}
-\eta\Delta u+\nabla p=0,& \hbox{in the wedge region},\\
\mathrm{div} u=0,&\hbox{in the wedge region},\\
u=-v_{ct}, & \hbox{on the bottom boundary},\\
u\cdot n=0, \partial_n u=0, & \hbox{on the upper boundary}.
\end{array}
\right.
\end{equation}
We choose a polar coordinate system as shown in \ref{fig:wedge0}.
Introduce a stream function $\psi$ for the Stokes equation. By the incompressibility condition, the stream
function can be written as (see e.g. in \cite{cox1986})
\begin{equation*}
\psi=r(a\sin\theta+b\cos\theta+c\theta\sin\theta+d\theta\cos\theta).
\end{equation*}
Then the velocity can be computed as
\begin{equation*}
u_{r}=-\frac{1}{r}\partial_{\theta}\psi,\qquad u_{\theta}=\partial_{r}\phi,
\end{equation*}
where $u_{r}$ and $u_{\theta}$ are the velocities in radial direction and in the angular direction, respectively.
The boundary condition in~\eqref{e:stokes} reads
\begin{equation*}
\left\{
\begin{array}{ll}
\partial_r\psi=0,\quad -\frac{1}{r} \partial_{\theta}\psi= v_{ct}, & \hbox{on } \theta=\theta_e,\\
\partial_r \psi=0,\quad \partial_{\theta\theta}\psi=0, & \hbox{on } \theta=0.\\
\end{array}
\right.
\end{equation*}
Direct computations using the form of $\psi$  give
\begin{equation*}
a=-\frac{v_{ct}\theta_e\cos\theta_e}{\theta_e-\sin\theta_e\cos\theta_e}, b=0, c=0, d=\frac{v_{ct}\sin\theta_e}{\theta_e-\sin\theta_e\cos\theta_e}.
\end{equation*}
This leads to
\begin{align*}
u_{r}&=\frac{v_{ct}}{r(\theta_e-\sin\theta_e\cos\theta_e)}\left((\theta_e\cos\theta_e-\sin\theta_e)\cos\theta+\sin\theta_e\theta\sin\theta\right),\\
u_{\theta}&=\frac{v_{ct}}{\theta_e-\sin\theta_e\cos\theta_e}\left(-\theta_e\cos\theta_e\sin\theta+\sin\theta_e\theta\cos\theta \right).
\end{align*}
Then the norm of the gradient of the velocity field can be computed as
\begin{equation*}
|\nabla u| =\frac{2}{r} \frac{v_{ct}\sin\theta_e\sin\theta}{\theta_e-\sin\theta_e\cos\theta_e}
\end{equation*}
We could calculate the viscous dissipation in the two dimensional wedge ($l_c<r<l$, $0<\theta<\theta_e$) as
\begin{equation}\label{e:dissWedge}
\Psi=\eta \int_{l_c}^{l}\int_{0}^{\theta_e} |\nabla u|^2 rd\theta dr=\frac{2\eta v_{ct}^2\sin\theta_e^2 \ln \varepsilon }{\theta_e-\sin\theta_e\cos\theta_e},
\end{equation}
where $\varepsilon={l}/{l_c}$.

{\color{black}We then compute the viscous dissipations in a bulk region with a circular liquid surface and a flat substrate. We need to solve the
Stokes equation~\eqref{e:stokes}
in  the region. In general, the equation can not be solved analytically since the velocity field is not axisymmetric even for a
circular liquid surface. Instead, we solve the equation numerically by a finite element method. A typical numerical
result for the velocity field is shown in Figure~\ref{fig:vel}, where the radius of the droplet is $1mm$ and the velocity
of the substrate is $0.182 m/s$. We can also compute the energy dissipation rates in the bulk region by neglecting some small
regions of size $50\mu m$ near the two contact points. Then we can calculate the ratio between the viscous dissipation in the bulk and
that in the wedge regions(given by \eqref{e:dissWedge}). The ratios are about $0.21, 0.22$ and $0.24$ for the cases
when $\theta_e=\pi/6$, $\pi/3$ and $\pi/2$, respectively.
 \begin{figure}[H]
  \centering
  \includegraphics[width=5.5in]{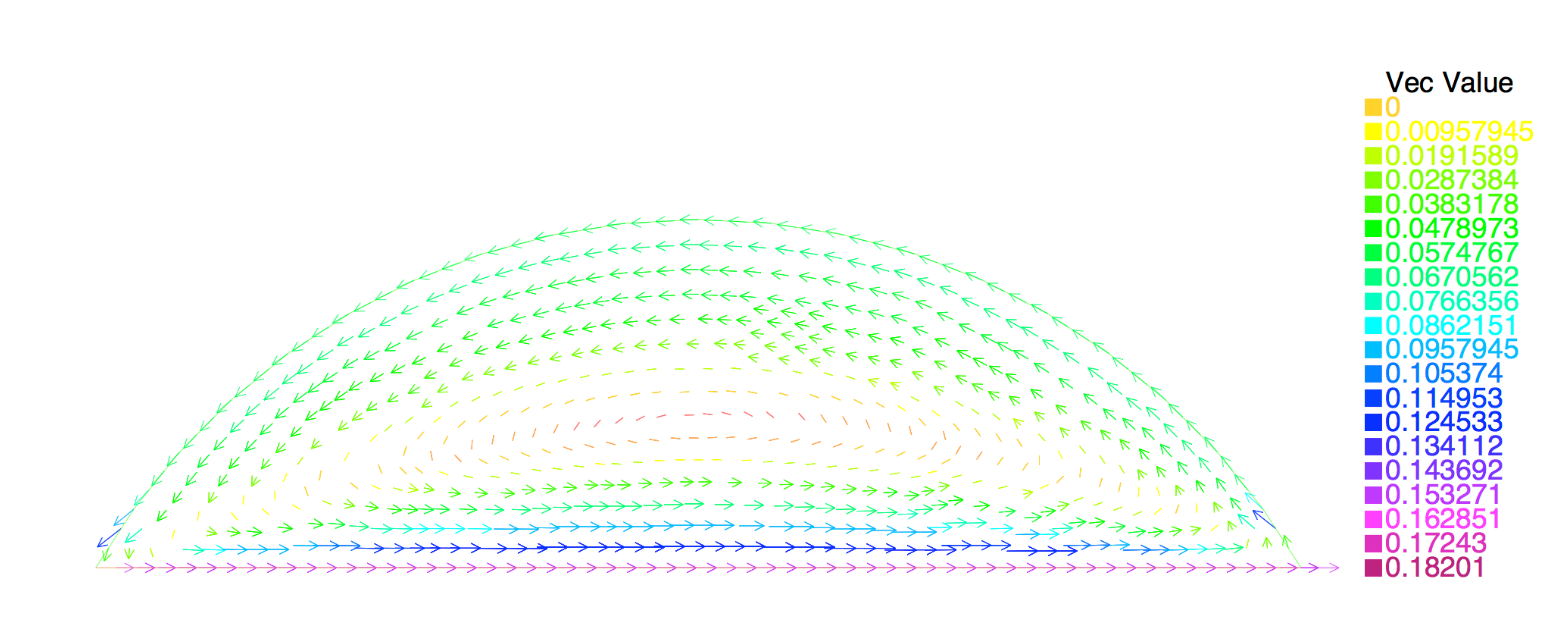}\\
  \caption{The velocity field in a droplet with contact angle $\theta_e=\pi/3$(with unit $m/s$).}\label{fig:vel}
\end{figure}
}

\subsection*{B. Asymptotic analysis}\label{AppendixB}
\setcounter{equation}{0}
\renewcommand{\theequation}{B\arabic{equation}}
In the appendix, we do asymptotic analysis for the equation~\eqref{gvs} under the condition that \addZero{$V_0|\kappa|^3\ll 1$}.
For simplicity in notation, we denote \addZero{ $a=\sin\alpha$ and }
$$\beta=\Big(\frac{3 V_0(-\kappa)^3}{\pi}\Big)^{\frac{1}{3}}=\Big(\frac{3 V_0}{\pi}\Big)^{\frac{1}{3}}(-\kappa)\ll 1.$$
Then the equation~\eqref{generalV0} is reduced
to
\begin{equation}\label{e:generalV1}
\beta^3 =  a^3\frac{(1-Y)(2+Y)}{X(1+Y)}- (2-3\sqrt{1-a^2}+(1-a^2)^\frac{3}{2}).
\end{equation}
One can verify that the parameter $a=\sin\alpha$ is also a small parameter in this case.
Therefore, by Taylor expansions with respect to the small parameter $a$, we easily see that
\begin{align*}
X &=\sqrt{1-a^2}\sin\theta_e+a\cos\theta_e= \sin\theta_e+a\cos\theta_e+O(a^2),\\
Y &=\sqrt{1-a^2}\cos\theta_e-a\sin\theta_e=  \cos\theta_e-a\sin\theta_e+O(a^2).
\end{align*}
We can also compute the following expansions (keep only the first two terms)
\begin{align*}
\frac{1}{1+Y}&\sim \frac{1}{1+\cos\theta_e} +\frac{\sin\theta_e}{(1+\cos\theta_e)^2} a,\\
\frac{1}{(1+Y)^2}&\sim \frac{1}{(1+\cos\theta_e)^2} +\frac{2\sin\theta_e}{(1+\cos\theta_e)^3} a,\\
\frac{2}{X(1+Y)} -\frac{Y}{X}&\sim Z_1+\frac{3}{(1+\cos\theta_e)^2}a.
\end{align*}
where $Z_1=\frac{1}{\sin\theta_e}\left(\frac{2}{1+\cos\theta_e}-\cos\theta_e\right)>0$.
We then do Taylor expansions for the right hand side term Eq.~\eqref{e:generalV1}
 and obtain
\begin{equation*}
\begin{aligned}
\beta^3
&= \left[\frac{2}{X(1+Y)}-\frac{Y}{X}\right]a^{3}-\left(2-3\sqrt{1-a^2}+(1-a^2)^{3/2}\right)\\
&\sim \left(Z_1+\frac{3a}{(1+\cos\theta_e)^2}\right)a^3+\left(2-3\left(1-\frac{a^2}{2}-\frac{a^4}{8}\right)+\left(1-\frac{3a^2}{2}+\frac{3a^4}{8}\right)\right)+O(a^4)\\
&\sim Z_1a^3+\frac{3}{4}Z_2 a^4,\\
\end{aligned}
\end{equation*}
where $Z_2=\frac{4}{(1+\cos\theta_e)^2}-1.$
 This leads to
\addTwo{
 \begin{equation}\label{e:as2}
 {a}\sim Z_1^{-\frac{1}{3}}\beta-\frac{Z_2Z_1^{-\frac{5}{3}}}{4}\beta^2=Z_1^{-\frac{1}{3}} \Big(\frac{3 V_0}{\pi}\Big)^{\frac{1}{3}}(-\kappa)
 -\frac{Z_2Z_1^{-\frac{5}{3}}}{4}\Big(\frac{3 V_0}{\pi}\Big)^{\frac{2}{3}}(-\kappa)^2,
 \end{equation}
 and
 }
 \begin{equation}\label{e:as}
 {a}\sim Z_1^{-\frac{1}{3}}\beta=Z_1^{-\frac{1}{3}} \Big(\frac{3 V_0}{\pi}\Big)^{\frac{1}{3}}(-\kappa),
 \end{equation}
 in leading order.

We then derive the asymptotic expansion of $\frac{\partial \mathcal{E}}{\partial \kappa}$ defined in Eq.~(\ref{generaldEdk}).
We first do Taylor expansions with respect to $a$ for the parameter A(defined in Eq.~\eqref{A}) to obtain,
\begin{equation}\label{e:Aexpn}
\begin{aligned}
A=&\left[\frac{2}{X(1+Y)}-\frac{Y}{X}\right]a(1-a^2)^{\frac{1}{2}}+\left[\frac{1}{(1+Y)^2}-1\right]a^2\\
&\!\!\!\!\!\!\sim Z_1a+Z_2a^2+O(a^3).
\end{aligned}
\end{equation}
Similarly, we can compute (keep only the first two terms)
\addTwo{
\begin{align}
\frac{-2\sqrt{1-a^2}}{1+Y}-\frac{aX}{(1+Y)^2}+\cos\theta_e
&\sim -Z_1\sin\theta_e-\frac{3\sin\theta_e}{(1+\cos\theta_e)^2}a,\\
\frac{a^2}{1+Y}-(1-\sqrt{1-a^2})\cos\theta_e& \sim \frac{Z_1}{2}\sin\theta_e a^2+\frac{\sin\theta_e}{(1+\cos\theta_e)^2} a^3.\label{e:Temp_expns}
\end{align}
We then approximate the Eq.~(\ref{generalV0}) according to Eq.~(\ref{e:Temp_expns}) and (\ref{e:as2})
\begin{equation}
\begin{aligned}
\label{appro_E}
\mathcal{E}&=2\pi (-\kappa)^{-2}\gamma \left[\displaystyle\frac{ a^2}{1+ Y }- (1-\sqrt{1-a^2})\cos \theta_e \right]\\
&\sim 2\pi(-\kappa)^{-2}\gamma\left[\frac{Z_1}{2}\sin{\theta_e}a^2+\frac{\sin\theta_e a^3}{(1+\cos\theta_e)^2}\right]\\
&\sim\gamma \sin{\theta_e}\left[(3V_0)^{\frac{2}{3}}(\pi Z_1)^{\frac{1}{3}} -\frac{3V_0 \kappa}{2Z_1}  \right].
\end{aligned}
\end{equation}
}
We then do direct calculations for  the right hand side term of  Eq.~(\ref{generaldEdk}),
\begin{equation}\label{e:engs}
\begin{aligned}
\frac{d \mathcal{E}}{d \kappa }
=&\frac{2\pi\gamma }{A(-\kappa)^{3}}\left\{\beta^3 \left[\frac{-2\sqrt{1-a^2}}{1+Y}-\frac{aX}{(1+Y)^2}+\cos\theta_e\right]+2 A \left[\frac{a^2}{1+Y}-(1-\sqrt{1-a^2})\cos\theta_e\right]\right\}\\
&\!\!\!\!\!\!\sim \frac{2\pi\gamma }{Z_1 a(-\kappa)^{3}}\left\{ (Z_1a^3+\frac{3Z_2 a^4}{4})\left[-Z_1\sin\theta_e-\frac{3\sin\theta_e}{(1+\cos\theta_e)^2}a\right]\right.\\
&\qquad\qquad \left.+2 (Z_1a+Z_2a^2) \left[\frac{Z_1}{2}\sin\theta_e a^2+\frac{\sin\theta_e}{(1+\cos\theta_e)^2} a^3\right]\right\}\\
&\!\!\!\!\!\!\sim \frac{2\pi\gamma\sin\theta_e a^3}{(-\kappa)^3}  \left[-\frac{1}{(1+\cos\theta_e)^2}+ \frac{Z_2}{4}\right]\\
&\!\!\!\!\!\!= 6\gamma V_0\sin\theta_eZ_1^{-1}\left[-\frac{1}{(1+\cos\theta_e)^2}+ \frac{Z_2}{4}\right]
= -\frac{3\gamma V_0\sin\theta_e}{2Z_1}.
\end{aligned}
\end{equation}
where in the second equation, we have used the above asymptotic expansions in \eqref{e:Aexpn} and \eqref{e:Temp_expns}; in the third equation, we keep only
the leading order term; and in the forth equation, we use Eq.~\eqref{e:as}.

Using the above formula for $\frac{d \mathcal{E}}{d\kappa}$, the equation~\eqref{gvs} is reduced to
\begin{equation}
\begin{aligned}
\dot{\vec{x}}&= \frac{\theta_e-\sin\theta_e \cos\theta_e}{2\eta\pi (-\kappa)^{-1}a|\ln\varepsilon|\sin^2\theta_e}\frac{3\gamma V_0\sin\theta_e}{2Z_1} \nabla_{\Gamma}\kappa,\\
&=\frac{\gamma (\theta_e-\sin\theta_e \cos\theta_e)}{4\eta |\ln\varepsilon|\sin \theta_e}\Big(\frac{3 V_0}{\pi Z_1}\Big)^{\frac{2}{3}}  \nabla_{\Gamma}\kappa,
\end{aligned}
\end{equation}
where we have used Eq.~\eqref{e:as} in the second equation. This gives the reduced equation~\eqref{e:reducedEq}.

\section*{References}

\bibliographystyle{plain}%
\bibliography{bibfile}

\begin{thebibliography}{10}

\bibitem{2020How}
R.~Bhardwaj and A.~Agrawal.
\newblock How coronavirus survives for days on surfaces.
\newblock {\em Physics of Fluids}, 32(11), 2020.

\bibitem{bjelobrk2016thermocapillary}
N.~Bjelobrk, H.~L. Girard, S.~P.~B. Subramanyam, H.~M. Kwon, D.~Qu{\'e}r{\'e},
  and K.~K. Varanasi.
\newblock Thermocapillary motion on lubricant-impregnated surfaces.
\newblock {\em Physical Review Fluids}, 1(6):063902, 2016.

\bibitem{bonn2009wetting}
D.~Bonn, J.~Eggers, J.~Indekeu, J.~Meunier, and E.~Rolley.
\newblock Wetting and spreading.
\newblock {\em Reviews of modern physics}, 81(2):739, 2009.

\bibitem{chan2020directional}
T.~S. Chan, F.~Yang, and A.~Carlson.
\newblock Directional spreading of a viscous droplet on a conical fibre.
\newblock {\em Journal of Fluid Mechanics}, 894, 2020.

\bibitem{chatterjee2017surface}
S.~Chatterjee, S.~Bhattacharjee, S.~Maurya, V.~Srinivasan, K.~Khare, and
  S.~Khandekar.
\newblock Surface wettability of an atomically heterogeneous system and the
  resulting intermolecular forces.
\newblock {\em EPL (Europhysics Letters)}, 118(6):68006, 2017.

\bibitem{chatterjee2019wetting}
S.~Chatterjee, Krishn~P. Singh, and S.~Bhattacharjee.
\newblock Wetting hysteresis of atomically heterogeneous systems created by low
  energy inert gas ion irradiation on metal surfaces: Liquid thin film coverage
  in the receding mode and surface interaction energies.
\newblock {\em Applied Surface Science}, 470:773--782, 2019.

\bibitem{chaudhury1992make}
M.~K. Chaudhury and G.~M. Whitesides.
\newblock How to make water run uphill.
\newblock {\em Science}, 256(5063):1539--1541, 1992.

\bibitem{chen2018ultrafast}
H.~Chen, T.~Ran, Y.~Gan, J.~Zhou, Y.~Zhang, L.~Zhang, D.~Zhang, and L.~Jiang.
\newblock Ultrafast water harvesting and transport in hierarchical
  microchannels.
\newblock {\em Nature Materials}, 17(10):935--942, 2018.

\bibitem{cox1986}
R.~G. Cox.
\newblock The dynamics of the spreading of liquids on a solid surface. {Part 1.
  Viscous flow}.
\newblock {\em Journal of Fluid Mechanics}, 168:169--194, 1986.

\bibitem{Gennes03}
P.G. de~Gennes, F.~Brochard-Wyart, and D.~Quere.
\newblock {\em Capillarity and Wetting Phenomena}.
\newblock Springer Berlin, 2003.

\bibitem{dervaux2020nonlinear}
J.~Dervaux, M.~Roch{\'e}, and L.~Limat.
\newblock Nonlinear theory of wetting on deformable substrates.
\newblock {\em Soft Matter}, 16(22):5157--5176, 2020.

\bibitem{doi2013soft}
M.~Doi.
\newblock {\em Soft matter physics}.
\newblock Oxford University Press, 2013.

\bibitem{Doi15}
M.~Doi.
\newblock Onsager priciple as a tool for approximation.
\newblock {\em Chinese Physics B}, 24:020505, 2015.

\bibitem{doi2019application}
M.~Doi, J.~Zhou, Y.~Di, and X.~Xu.
\newblock Application of the onsager-machlup integral in solving dynamic
  equations in nonequilibrium systems.
\newblock {\em Physical Review E}, 99(6):063303, 2019.

\bibitem{dos1995free}
F.~D. Dos~Santos and T.~Ondarcuhu.
\newblock Free-running droplets.
\newblock {\em Physical Review Letters}, 75(16):2972, 1995.

\bibitem{duprat2012wetting}
C.~Duprat, S.~Protiere, A.~Y. Beebe, and H.A. Stone.
\newblock Wetting of flexible fibre arrays.
\newblock {\em Nature}, 482(7386):510--513, 2012.

\bibitem{extrand1995liquid}
C.~W. Extrand and Y.~Kumagai.
\newblock Liquid drops on an inclined plane: the relation between contact
  angles, drop shape, and retentive force.
\newblock {\em Journal of colloid and interface science}, 170(2):515--521,
  1995.

\bibitem{galatola2018spontaneous}
P.~Galatola.
\newblock Spontaneous capillary propulsion of liquid droplets on substrates
  with nonuniform curvature.
\newblock {\em Physical Review Fluids}, 3(10):103601, 2018.

\bibitem{giacomello2016wetting}
A.~Giacomello, L.~Schimmele, and S.~Dietrich.
\newblock Wetting hysteresis induced by nanodefects.
\newblock {\em Proceedings of the National Academy of Sciences},
  113(3):E262--E271, 2016.

\bibitem{guo2013direct}
S.~Guo, M.~Gao, X.~Xiong, Y.~Wang, X.~Wang, P.~Sheng, and P.~Tong.
\newblock Direct measurement of friction of a fluctuating contact line.
\newblock {\em Physical Review Letters}, 111(2):026101, 2013.

\bibitem{guo2019onset}
S.~Guo, X.~Xu, T.~Qian, Y.~Di, M.~Doi, and P.~Tong.
\newblock Onset of thin film meniscus along a fibre.
\newblock {\em Journal of Fluid Mechanics}, 865:650--680, 2019.

\bibitem{1992Intermolecular}
J.~N. Israelachvili.
\newblock Intermolecular and surface forces.
\newblock {\em AcademicPress}, 1992.

\bibitem{jiang2018solid}
W.~Jiang, Y.~Wang, D.~J. Srolovitz, and W.~Bao.
\newblock Solid-state dewetting on curved substrates.
\newblock {\em Physical Review Materials}, 2(11):113401, 2018.

\bibitem{joanny1984model}
J.F. Joanny and P.-G. De~Gennes.
\newblock A model for contact angle hysteresis.
\newblock {\em The journal of chemical physics}, 81(1):552--562, 1984.

\bibitem{johnson1964contact}
R.~E. Johnson~Jr and R.~H. Dettre.
\newblock Contact angle hysteresis. iii. study of an idealized heterogeneous
  surface.
\newblock {\em The journal of physical chemistry}, 68(7):1744--1750, 1964.

\bibitem{Ju2012A}
J.~Ju, H.~Bai, Y.~Zheng, T.~Zhao, R.~Fang, and L.~Jiang.
\newblock A multi-structural and multi-functional integrated fog collection
  system in cactus.
\newblock {\em Nature Communications}, 3:1247, 2012.

\bibitem{keiser2018dynamics}
L.~Keiser, K.~Jaafar, J.~Bico, and E.~Reyssat.
\newblock Dynamics of non-wetting drops confined in a hele-shaw cell.
\newblock {\em Journal of Fluid Mechanics}, 845:245--262, 2018.

\bibitem{lam2001effect}
C.~Lam, N.~Kim, D.~Hui, D.~Kwok, M.L. Hair, and A.W. Neumann.
\newblock The effect of liquid properties to contact angle hysteresis.
\newblock {\em Colloids and Surfaces A: Physicochemical and Engineering
  Aspects}, 189(1-3):265--278, 2001.

\bibitem{lam2002study}
C.~Lam, R.~Wu, D.~Li, M.L. Hair, and A.W. Neumann.
\newblock Study of the advancing and receding contact angles: liquid sorption
  as a cause of contact angle hysteresis.
\newblock {\em Advances in colloid and interface science}, 96(1-3):169--191,
  2002.

\bibitem{li2017external}
Y.~Li, L.~He, X.~Zhang, N.~Zhang, and D.~Tian.
\newblock External-field-induced gradient wetting for controllable liquid
  transport: From movement on the surface to penetration into the surface.
\newblock {\em Advanced Materials}, 29(45):1703802, 2017.

\bibitem{liu2015effective}
C.~Liu, Y.~Xue, Y.~Chen, and Y.~Zheng.
\newblock Effective directional self-gathering of drops on spine of cactus with
  splayed capillary arrays.
\newblock {\em Scientific reports}, 5(1):1--8, 2015.

\bibitem{lorenceau2004drops}
{E}. Lorenceau and D.~Qu{\'e}r{\'e}.
\newblock Drops on a conical wire.
\newblock {\em Journal of Fluid Mechanics}, 510:29, 2004.

\bibitem{LvSubstrate}
C.~Lv, C.~Chen, Y.~Chuang, F.~Tseng, Y.~Yin, F.~Grey, and Q.~Zheng.
\newblock Substrate curvature gradient drives rapid droplet motion.
\newblock {\em Physical Review Letters}, 113(2):026101, 2014.

\bibitem{lv2021wetting}
C.~Lv and S.~Hardt.
\newblock Wetting of a liquid annulus in a capillary tube.
\newblock {\em Soft Matter}, 17(7):1756--1772, 2021.

\bibitem{ManDoi2016}
X.~Man and M.~Doi.
\newblock Ring to mountain transition in deposition pattern of drying droplets.
\newblock {\em Physical Review Letters}, 116(6):066101, 2016.

\bibitem{marmur2017contact}
A.~Marmur, C.~Della~Volpe, S.~Siboni, A.~Amirfazli, and J.~W. Drelich.
\newblock Contact angles and wettability: towards common and accurate
  terminology.
\newblock {\em Surface Innovations}, 5(1):3--8, 2017.

\bibitem{mayo2015simulating}
L.~C. Mayo, S.~W. McCue, T.~J. Moroney, A.~Forster, W, D.~M. Kempthorne, J.~A.
  Belward, and I.~W. Turner.
\newblock Simulating droplet motion on virtual leaf surfaces.
\newblock {\em Royal Society open science}, 2(5):140528, 2015.

\bibitem{JohnDynamics}
J.~McCarthy, D.~Vella, and A.~A. Castrej{\'o}n-Pita.
\newblock Dynamics of droplets on cones: self-propulsion due to curvature
  gradients.
\newblock {\em Soft Matter}, 15:9997, 2019.

\bibitem{onsager1931phys}
L.~Onsager.
\newblock Reciprocal relations in irreversible processes. i.
\newblock {\em Physical Reviews}, 37:405--426, 1931.

\bibitem{onsager1931phys1}
L.~Onsager.
\newblock Reciprocal relations in irreversible processes. ii.
\newblock {\em Physical Reviews}, 38:2265--2279, 1931.

\bibitem{prakash2008surface}
M.~Prakash, D.~Qu{\'e}r{\'e}, and John W.~M. Bush.
\newblock Surface tension transport of prey by feeding shorebirds: the
  capillary ratchet.
\newblock {\em Science}, 320(5878):931--934, 2008.

\bibitem{renvoise2009drop}
P.~Renvois{\'e}, J.~W.~M. Bush, M.~Prakash, and D.~Qu{\'e}r{\'e}.
\newblock Drop propulsion in tapered tubes.
\newblock {\em EPL (Europhysics Letters)}, 86(6):64003, 2009.

\bibitem{reyssat2014drops}
E.~Reyssat.
\newblock Drops and bubbles in wedges.
\newblock {\em Journal of fluid mechanics}, 748:641--662, 2014.

\bibitem{sumino2005chemosensitive}
Y.~Sumino, H.~Kitahata, K.~Yoshikawa, M.~Nagayama, N.~Magome, and Y.~Mori.
\newblock Chemosensitive running droplet.
\newblock {\em Physical Review E}, 72(4):041603, 2005.

\bibitem{sumino2005self}
Y.~Sumino, N.~Magome, T.~Hamada, and K.~Yoshikawa.
\newblock Self-running droplet: Emergence of regular motion from nonequilibrium
  noise.
\newblock {\em Physical Review Letters}, 94(6):068301, 2005.

\bibitem{tanner1979spreading}
L.H. Tanner.
\newblock The spreading of silicone oil drops on horizontal surfaces.
\newblock {\em Journal of Physics D: Applied Physics}, 12(9):1473, 1979.

\bibitem{Qianbin2015Self}
Q.~Wang, X.~Yao, H.~Liu, D.~Qu{\'e}r{\'e}, and L.~Jiang.
\newblock Self-removal of condensed water on the legs of water striders.
\newblock {\em Proceedings of the National Academy of Sciences},
  112:9247--9252.

\bibitem{wang2020enhancing}
Z.~Wang, A.~Owais, C.~Neto, J.~Pereira, and Y.~Gan.
\newblock Enhancing spontaneous droplet motion on structured surfaces with
  tailored wedge design.
\newblock {\em Advanced Materials Interfaces}, 8(2):2000520, 2020.

\bibitem{xu2016variational}
X.~Xu, Y.~Di, and M.~Doi.
\newblock Variational method for contact line problems in sliding liquids.
\newblock {\em Physics of Fluids}, 28(8):087101, 2016.

\bibitem{xu2020theoretical}
X.~Xu, M.~Doi, J.~Zhou, and Y.~Di.
\newblock Theoretical analysis for flattening of a rising bubble in a
  hele--shaw cell.
\newblock {\em Physics of Fluids}, 32(9):092102, 2020.

\bibitem{xu2012thermal}
X.~Xu and T.~Qian.
\newblock Thermal singularity and droplet motion in one-component fluids on
  solid substrates with thermal gradients.
\newblock {\em Physical Review E}, 85(6):061603, 2012.

\bibitem{XuWang2020}
X.~Xu and X.~Wang.
\newblock Theoretical analysis for dynamic contact angle hysteresis on
  chemically patterned surfaces.
\newblock {\em Physics of Fluids}, 2020.

\bibitem{xu2012droplet}
X.-P. Xu and T.~Qian.
\newblock Droplet motion in one-component fluids on solid substrates with
  wettability gradients.
\newblock {\em Physical Review E}, 85(5):051601, 2012.

\bibitem{yang2008conversion}
J.~Yang, Z.~Yang, C.~Chen, and D.~Yao.
\newblock Conversion of surface energy and manipulation of a single droplet
  across micropatterned surfaces.
\newblock {\em Langmuir}, 24(17):9889--9897, 2008.

\bibitem{zheng2010directional}
Y.~Zheng, H.~Bai, Z.~Huang, X.~Tian, F.~Q. Nie, Y.~Zhao, J.~Zhai, and L.~Jiang.
\newblock Directional water collection on wetted spider silk.
\newblock {\em Nature}, 463(7281):640--643, 2010.

\end{thebibliography}

\end{document}